\documentclass[twocolumn]{aastex62}
\usepackage{amsmath}
\usepackage[normalem]{ulem}

\graphicspath{{./}{figures/}}
\begin{document}

\title{Modeling Dust Production, Growth, and Destruction in Reionization-Era Galaxies with the CROC Simulations: Methods and Parameter Exploration}

\correspondingauthor{Clarke Esmerian}
\email{cesmerian@uchicago.edu}

\author{Clarke J.\ Esmerian}
\affiliation{Department of Astronomy \& Astrophysics, University of Chicago\\
Chicago, IL 60637 USA}
\affiliation{Kavli Institute for Cosmological Physics;
The University of Chicago;
Chicago, IL 60637 USA}

\author{Nickolay Y.\ Gnedin}
\affiliation{Fermi National Accelerator Laboratory;
Batavia, IL 60510, USA}
\affiliation{Kavli Institute for Cosmological Physics;
The University of Chicago;
Chicago, IL 60637 USA}
\affiliation{Department of Astronomy \& Astrophysics, University of Chicago\\
Chicago, IL 60637 USA}

\begin{abstract}
We introduce a model for the explicit evolution of interstellar dust in a cosmological galaxy formation simulation. We post-process a simulation from the Cosmic Reionization on Computers project \citep[CROC,][]{Gnedin2014}, integrating an ordinary differential equation for the evolution of the dust-to-gas ratio along pathlines in the simulation sampled with a tracer particle technique. This model incorporates the effects of dust grain production in asymptotic giant branch (AGB) star winds and supernovae (SN), grain growth due to the accretion of heavy elements from the gas phase of the interstellar medium (ISM), and grain destruction due to thermal sputtering in the high temperature gas of supernova remnants (SNRs). A main conclusion of our analysis is the importance of a carefully chosen dust destruction model, for which different reasonable parameterizations can predict very different values at the $\sim 100$ pc resolution of the ISM in our simulations. We run this dust model on the single most massive galaxy in a 10$h^{-1}$ co-moving Mpc box, which attains a stellar mass of $\sim 2\times10^9 M_{\odot}$ by $z=5$. We find that the model is capable of reproducing dust masses and dust-sensitive observable quantities broadly consistent with existing data from high-redshift galaxies. The total dust mass in the simulated galaxy is somewhat sensitive to parameter choices for the dust model, especially the timescale for grain growth due to accretion in the ISM. Consequently, observable quantities that can constrain galaxy dust masses at these epochs are potentially useful for placing constraints on dust physics.
\end{abstract}

\keywords{dust -- galaxies: formation -- cosmology: theory -- methods: numerical}

\section{Introduction}

Cosmic dust is the solid phase of interstellar matter. While it constitutes at most a few percent of the mass in the Inter-Stellar Medium (ISM) of galaxies \citep[e.g.][]{RemyRuyer2014,Aniano2020,Galliano2021}, it is an important source of opacity in the ISM \citep{DraineBook}, and can thereby have a significant impact on the observable properties of galaxies \citep[e.g.][]{Galliano2018, Zavala2021}. A complete physical explanation for the observed properties of the galaxy population throughout cosmic time must therefore account for the dust content -- at the very least the amount and spatial distribution -- in individual galaxies. 

Such an explanation is still incomplete in large part because dust grains are subject to complicated physical processes in a variety of interstellar environments \citep[see][for a review]{Draine2003}. They are observed to be produced in the winds of evolved stars \citep{Lagadec2008, Srinivasan2009, Matsuura2009, Riebel2012, Nanni2019, HofnerOlofsson2018} and the remnants of supernova explosions \citep{Dunne2003,Gomez2012,DeLooze2017,Matsuura2011,SarangiMatsuuraMicelotta2018, MichelottaMatsuuraSarangi2018}. They may grow through the accretion of heavy elements in the gas phase of the ISM \citep{Draine1990,Dwek1998,WeingartnerDraine1999,MattssonAndersen2012, Feldmann2015}. They may also be destroyed by various physical processes that dissociate grains such as sputtering and sublimation \citep{DraineSalpeter1979a, DraineSalpeter1979b, Jones2004, Hoang2020}. Consequently, the prediction of the dust content in galaxies requires an accounting for these physical processes -- about all of which there remain deep theoretical uncertainties -- in a dynamical model of the ISM. 

Cosmological, fluid-dynamical galaxy formation simulations can provide such a model. By employing numerical methods that adaptively refine the simulation resolution in the areas of highest density, these simulations can resolve individual galaxy disk scale lengths ($\sim 1$kpc) in cosmologicaly representative volumes ($\gtrsim 10$Mpc, e.g.), or can resolve the disk scale height ($\sim 100$pc) for the entire cosmological Lagrangian volume of an individual galaxy. Consequently, these models have the ability to simultaneously model the dynamics of the ISM (albeit with a physical fidelity inherently limited by resolution) while self-consistently accounting for the global environment of the galaxy, including its history of star formation and metal enrichment over cosmic time. These simulations have recently achieved success in reproducing many observed galaxy and ISM scaling relations \citep[see e.g.][for recent reviews of galaxy formation simulations]{SomervilleDave2015, NaabOstriker2017, FaucherGiguere2018, Vogelsberger2020}

The explicit coupling of interstellar metal enrichment to the predicted star formation history is commonplace -- simulations predict the locations and properties of stellar populations formed from local ISM condistions, and account for the energetic and chemical feedback of these populations on the surrounding ISM based on the tabulated predictions of stellar evolution models \citep[e.g.][]{Leitherer1999,ConroyGunn2010}. Metals are then typically assumed to be returned to the gas phase of the ISM and passively advected, modifying the local radiative cooling rates accordingly. 

While the dust content of such simulations can be estimated to zeroth-order by assuming a constant dust-to-metal ratio, recent efforts have begun to include an explicit treatment for the creation, growth, and destruction of dust as a function of local ISM conditions, thereby enabling a prediction of the spatially and temporally varying dust-to-metal ratio  self-consistent with the properties of the simulated ISM \citep[e.g.][]{Bekki2015, McKinnon2016, McKinnon2017, Aoyama2018, Gjergo2018, Hou2019, Vogelsberger2019, Li2019, Graziani2020, Li2021, Granato2021, Parente2022, Trebitsch2021, Kannan2022, Choban2022}. These efforts have already provided insights into the physical processes governing dust evolution in galaxies throughout cosmic time. However, they must necessarily contend with the still profound uncertainties in the physics governing dust in the ISM. 

The complexity of these processes render any attempt at first-principle estimates of their rates and dependence on ISM properties order-of-magnitude at best. Consequently, the equations governing the evolution of the dust-to-metal ratio include a number of free parameters used to quantify these uncertainties, which must be systematically varied to understand their effect on the predicted dust content and compared with observation. The computational expense of running fully coupled galaxy formation and dust physics simulations severely prohibits the size of this parameter space that can be feasibly explored. 

This difficulty is only be compounded by the increasingly stringent observational constraints on the dust content of galaxies, and this progress is certain to accelerate in the near future. Already the dust content of galaxies in the local universe has been well characterized \citep[see][for a recent review]{Galliano2018}, and knowledge of the earlier universe is expanding rapidly \citep[e.g.][for a recent review]{PerouxHowk2020}. The explosion of data anticipated from JWST will enable a census of the cosmic dust content to high redshift, demanding unprecedented precision from theoretical modeling.

Already, existing observational constraints are suggesting the richness of data to come. There are a growing number of galaxies at $z \gtrsim 6$ for which there are observational indications of substantial dust content \citep[e.g.][]{Bertoldi2003, Venemans2012, Willott2013, Wang2013, Riechers2013, Cooray2014, Watson2015, daCunha2015, Dessauges-Zavadsky2017, Knudsen2017, Laporte2017, Decarli2017, Venemans2017, Strandet2017, Izumi2018, Marrone2018, Hashimoto2019, Tamura2019, Bakx2020, Fudamoto2021arXiv}, while there are also similarly many results showing little or no dust in galaxies of the same epoch \citep[e.g.][]{Walter2012, Bouwens2012, Kanekar2013, Ouchi2013, Capak2015, Maiolino2015, Schaerer2015, Knudsen2016,Bradac2017,Matthee2019}. Together, these observations indicate a diversity in the dust content of galaxies in the early universe. Since the timescales regulating this dust content may be comparable to the age of the universe at this epoch, this diversity suggests promise for the ability of high-redshift observations to place unique constraints on the physics of dust in the ISM.

To enable physical interpretation of these existing and future constraints, in this paper we introduce a model for the explicit evolution of interstellar dust in a cosmological galaxy formation simulation. Specifically, we post-process a simulation from the Cosmic Reionization on Computers project \citep[CROC,][]{Gnedin2014}, integrating an ordinary differential equation (ODE) for the evolution of the dust-to-gas ratio along pathlines in the simulation sampled with a tracer particle technique. This model incorporates the effects of dust grain production in asymptotic giant branch (AGB) star winds and supernovae (SN), grain growth in the ISM due to accretion of heavy elements from the gas phase, and grain destruction due to thermal sputtering in high temperature gas of supernova remnants (SNRs). This allows us to predict the fraction of metals in the ISM that are locked up in dust grains as a function of space and time in the simulation. Using tracer particles in post-processing enables us to run many models and fully explore the parameter space of deeply uncertain dust physics processes. Consequently, this allows us to compare dust-dependent observable quantities in the ISM of high-redshift galaxies with data, and predict the dependence of these predictions on the assumed dust physical model.

The structure of the paper is as follows: in Section 2 we review the methods of the CROC simulations, present the methods of our model - both the equations that constitute the physical model and the numerical methods used to solve them, and describe our procedure for predicting observational quantities from our model results. In Section 3 we present the preliminary results of the model applied to a simulated massive galaxy ($M_{\star} \approx 2\times10^9M_{\odot}$ by $z=5$) -- both the predictions of its dust content for a wide range of dust model parameters, and the comparisons of these predictions to observational data, as a proof-of-concept. In Section 4 we discuss the implications of these results as well as uncertainties in our modeling, comparing to previous efforts. In Section 5 we conclude. We will present the statistically meaningful results of a larger set of simulations in a companion paper. 

\section{Methods} 

\begin{figure}
    \centering
    \includegraphics[width=\linewidth]{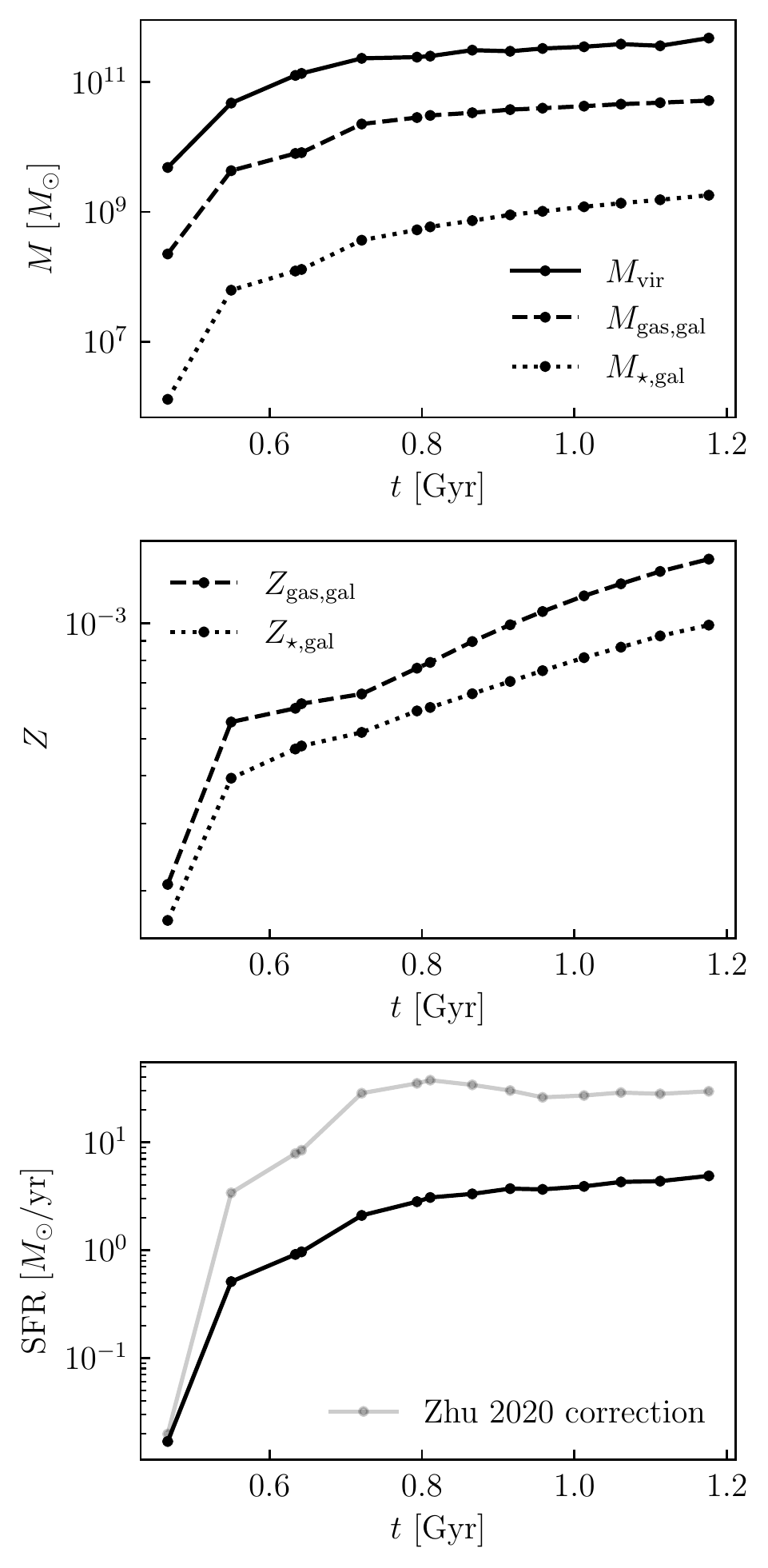}
    \caption{Simulated galaxy masses, metallicities, and star formation rate as a function of cosmological time. The top panel shows the evolution of the total virial mass, galaxy stellar mass, and galaxy gas mass of the most massive halo in the simulation. The middle panel shows the mass-weighted average galaxy gas and stellar metallicities. The bottom pannel shows the star formation rate of the central galaxy, and the same rate corrected by the factor determined by \citet{Zhu2020} to correct stellar masses to observations, stated in Eq.~\ref{eq:SFR_corr}.}
    \label{fig:galaxy_properties}
\end{figure}

\subsection{Simulation: Cosmic Reionization on Computers}

To model the ISM of high-redshift galaxies, we employ the Cosmic Reionization on Computers (CROC) numerical galaxy formation simulations described in \citet{Gnedin2014,GnedinKaurov2014,Gnedin2016}, to which we refer the reader for details. Here we summarize the main components of the simulation machinery.

\subsubsection{CROC Model}

CROC uses the Adaptive Refinement Tree (ART) code \citep{Kravtsov1999, Kravtsov2002, Rudd2008} to solve the three-dimensional inviscid fluid equations coupled gravitationally to collisionless cold dark matter with cosmological initial and boundary conditions. Cooling and heating processes are accounted for as tabulated in \citet{GnedinHollon2012}. Radiative transfer is fully coupled to the gas dynamics and chemistry and solved using the Optically Thin Variable Edington Tensor (OTVET) approximation \citep{GnedinAbel2001, Gnedin2014}. Molecular hydrogen fractions are calculated from local ISM conditions using the fitting functions of \citet{GnedinDraine2014}. Star formation occurs in molecular gas with a fixed depletion time of $\tau_{\rm SF} = 1.5$Gyr. Star particles are treated as single-age stellar populations with a \citet{Kroupa2001} IMF. Stellar feedback effects are incorporated with the delayed cooling subgrid model, adopting a delay timescale of $\tau_{\rm BW} = 10$Myr. Star particles source ionizing radiation with an escape fraction $\epsilon_{\rm UV} = 0.15$. 

The specific simulation we ran for this analysis is a cosmological volume $10h^{-1}$ co-moving Mpc on each side, assuming a flat $\Lambda$CDM cosmology with $\Omega_{\rm M} = 0.3036$, $\Omega_{\rm b} = 0.0479$, $\Omega_{\Lambda} = 0.6964$, and $h = H_0/(100{\rm km}\;{\rm s}^{-1}{\rm Mpc}^{-1}) = 0.6814$. An overdensity on the scale of the simulation box (i.e. the ``DC'' mode) of $\delta_{\rm DC} = 0.700869$ was accounted for using supercomoving variables as described in \citet{Gnedin2011}. Adaptive refinement is allowed down to a minimum cell size of $100$pc in physical units.

Halo identification is performed with the ROCKSTAR halo finder \citep{Behroozi2013}. We adopt the virial radius definition corresponding to a redshift-dependent over-density with respect to the critical density defined \citet{BryanNorman1998}. We find that during major mergers the halo center identified by ROCKSTAR imperfectly matches the location of the central galaxy, so we re-define the galaxy center as the median of stellar particle positions along each axis, using only stellar particles within the ROCKSTAR identified virial radius -- i.e. the median stellar particle x position, y position, and z position. Note that this does not necessarily correspond to the location of an actual star particle -- since the median particle position along each axis is not guaranteed to belong to the same particle -- but we have confirmed visually that this method provides a good estimate of the central galaxy position. We consider all stellar particles and gas cells within 0.1$R_{\rm vir}$ of this center as belonging to the central galaxy.

\subsubsection{Halo and Galaxy Properties}

Figure~\ref{fig:galaxy_properties} shows the evolution of basic simulated galaxy quantities. By the final snapshot of the simulation at $t=1.18$Gyr (corresponding to $z=5$), the galaxy resides in a dark matter halo of total mass $4.80\times10^{11}M_{\odot}$ with a stellar mass of $1.81\times10^9M_{\odot}$ and a gas mass of $5.22\times10^{10}M_{\odot}$. At this time the central galaxy has a gas and stellar metallicity $Z \approx 10^{-3} = 0.07 Z_{\odot}$, and a star formation rate (averaged over $10$Myr) of $4.88M_{\odot}/{\rm yr}$. 

We focus exclusively on the ISM of this galaxy for the remainder of our analysis.

\subsubsection{ISM Phase Structure}\label{subsubsec:ISM_phase_structure}

\begin{figure}
    \centering
    \includegraphics[width=\linewidth]{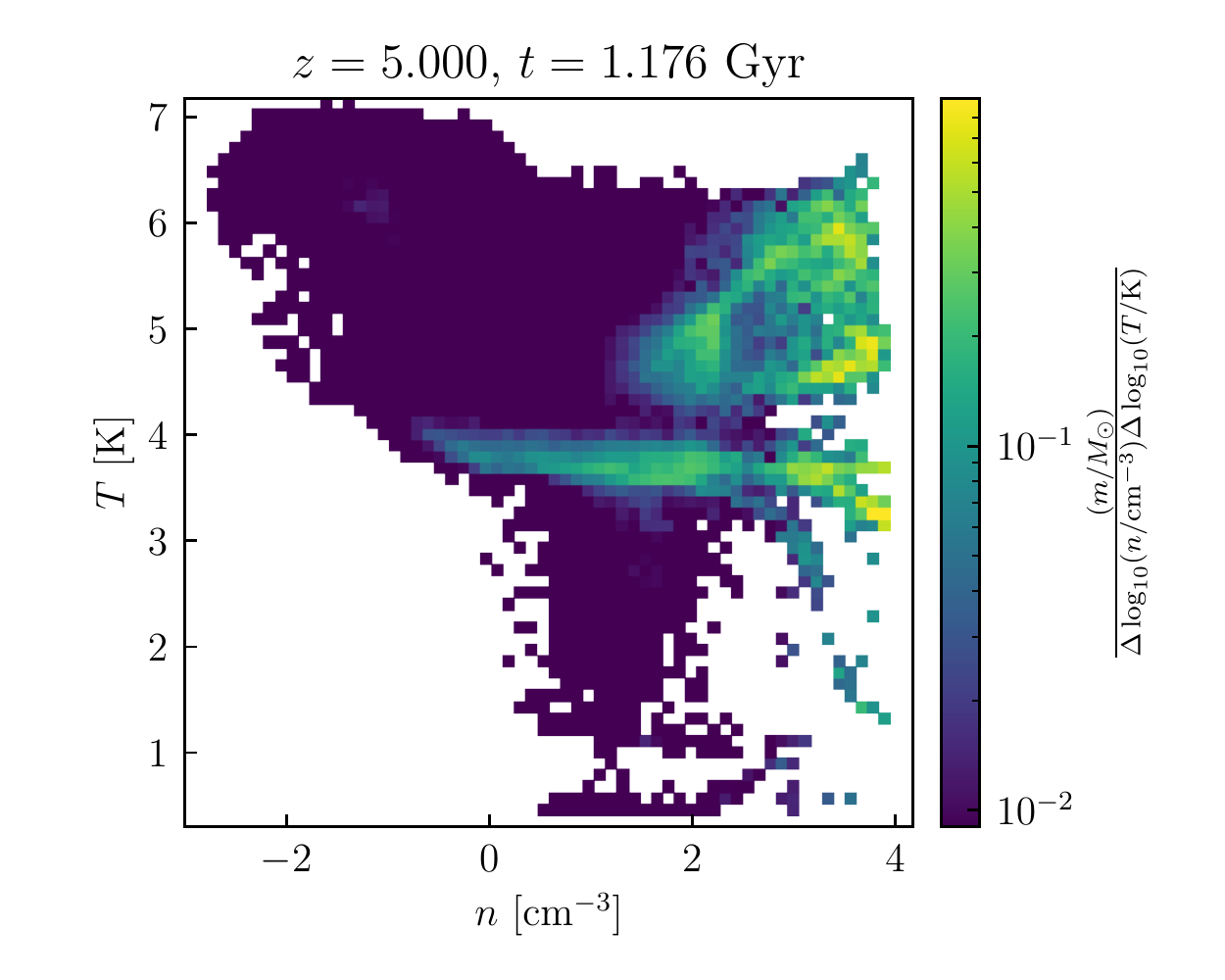}
    \caption{Temperature-density distribution the ISM of our simulated galaxy at $z=5$.}
    \label{fig:n_T_dist}
\end{figure}

\begin{figure}
    \centering
    \includegraphics[width=\linewidth]{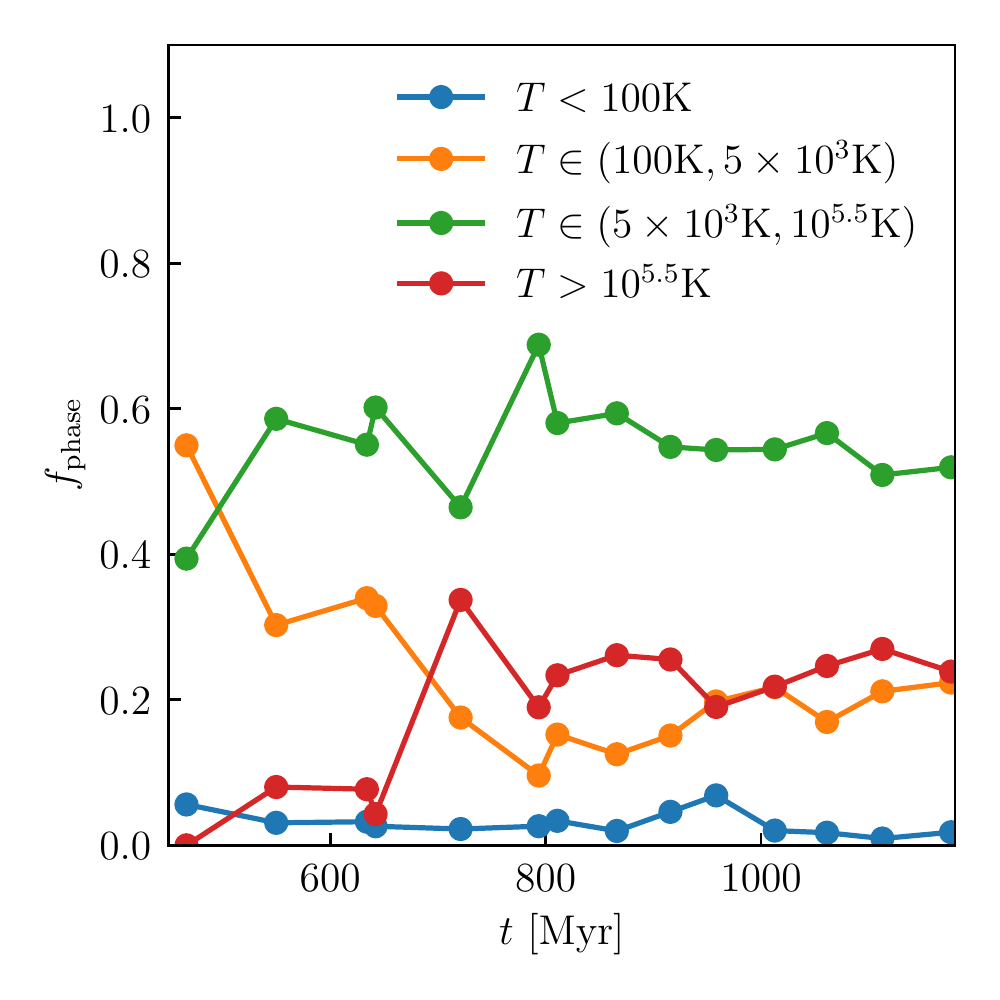}
    \caption{Mass fractions of ISM phases defined by temperature cuts as a function of cosmological time.}
    \label{fig:f_phase}
\end{figure}

Since dust growth and destruction terms depend on the local properties of the ISM, we first examine the distributions of these properties in the simulation. Figure~\ref{fig:n_T_dist} shows the temperature-density distribution of the ISM in the final snapshot of the simulation at $z=5$. At the resolution of this simulation (fixed to 100pc in proper units), the phase structure of the ISM does not appear to be well-resolved, with a substantial fraction of the ISM at temperatures ($T\sim 10^5$K) and densities ($n\sim10^3{\rm cm}^{-3}$). We quantify this in Figure~\ref{fig:f_phase}, which shows the mass fraction of four phases defined by temperature cuts: Cold ($T< 100$K), Warm Neutral Medium (WNM: $100{\rm K} < T< 5000{\rm K}$), Warm Ionized Medium (WIM: $5000{\rm K} < T < 10^{5.5}{\rm K}$ and Hot Ionized Medium (HIM: $T > 10^{5.5}$K). The consistent with Figure~\ref{fig:n_T_dist}, the mass fractions of the WIM and HIM are large (and the Cold and WNM are therefore small) compared to typical values expected from observations of the Milky Way \citep{DraineBook}. 

There is no a priori reason to expect galaxies located in Milky-Way mass halos at $z\gtrsim 6$ to have the ISM closely resembling the Milky Way one, and the mass fractions in various ISM phases in high redshift galaxies are not known. However, we are also aware that the CROC adopted stellar feedback recipe based on the delayed cooling algorithm results in model galaxies that do not match observations \citep{Zhu2020}. Hence, we are unable to demonstrate that the ISM structure in our model galaxies is correct. It is likely, therefore, that the high mass fraction of HIM is a numerical artifact from the inadequate stellar feedback recipe.

Furthermore, the blue line on the temperature panel of Figure~\ref{fig:tracer_average_histories} shows the molecular-fraction-weighted temperature of ISM tracers in the last few hundred million years of the simulation. The average temperature of molecular gas is predicted to be unrealistically high, $\sim 4000-6000$K. This is fully expected, as the temperature distribution of the ISM in this simulated galaxy is not sufficiently resolved with the mere 100pc resolution to be physically realistic. Consequently, in the default dust model, we do not use the local ISM temperature in the simulation to calculate dust growth and destruction rates. For the growth rate, which has a $\propto\sqrt{T}$ dependence (see Eq.~\ref{eq:dotD_acc}), we assume a constant $T = T_{\rm cold} = 50{\rm K}$ by default. For destruction, we calculate the rate based only on the assumption of a fixed dust mass destroyed per each (unresolved) supernova remnant, i.e. $D_{\rm dest} = D_{\rm SNR}$, which is independent of temperature (see Eq.~\ref{eq:dotD_acc}). As we show in the next section, calculating the destruction rate based on thermal sputtering (which depends on the local temperature) dramatically over-predicts the dust destruction rate, even when the growth rate also uses the local temperature instead of some fixed $T_{\rm cold}$. It will be interesting in the future to apply a model like this to higher resolution simulations which more realistically capture the local temperature distribution of the ISM and compare the destruction rates predicted by thermal sputtering compared to the unresolved SNR prescription we use here. 

\subsection{Pathline Sampling: Monte-Carlo Tracers}\label{subsec:methods_tracers}

As we detail in the next section, our dust model requires the assumption of several free parameters that quantify uncertainties in the physical processes that regulate dust evolution in the ISM. Since first-principles physics and observations generally do not constrain these parameters more tightly than within a few orders-of-magnitude, and the processes dependent on these parameters may interact in a complicated manner, it is necessary to explore a range of parameter values. Doing so would be prohibitively costly if each parameter combination required re-running the entire cosmological fluid-dynamical galaxy formation simulation. Therefore, when running the simulation we produce $\approx 10^4$ pathlines that sample the evolution of the simulated galaxy in an approximately Lagrangian manner. We can then post-process the simulation by integrating our dust model along each of these pathlines for as many different versions of the model that we want to explore without re-running the simulation. This dramatically reduces the computational cost and consequently greatly expands the size of the parameter grid we are able to explore. 

To calculate the pathlines, we use the Monte Carlo (hereafter MC) tracer method introduced in \citet{Genel2013}, to which we refer the reader for details. The tracer particles are initialized to randomly and uniformly sample the Lagrangian region for the dark matter halo of interest. Tracer particle positions are output at $\sim 5000$ snapshots for an average interval of $0.2$ Myr between each. At every snapshot, the tracers samples the local gas number density $n$, temperature $T$, metallicity $Z$, neutral hydrogen fraction $f_{\rm HI}$, molecular hydrogen fraction $f_{\rm H2}$, the metal production rate from supernovae (SN) $\dot{Z}_{\rm SN}$, the metal production from asymptotic giant branch (AGB) stars $\dot{Z}_{\rm AGB}$, and the rate density of supernovae $\dot{n}_{\rm SN}$. 
The quantities $\dot{Z}_{\rm SN}$, $\dot{Z}_{\rm AGB}$, and $\dot{n}_{\rm SN}$ are assigned from the stellar particles cells using Nearest Grid Point (NGP) interpolation. This makes the sampled quantities somewhat more noisy and avoids extra smoothing, but the MC tracers are already noisy by construction. Note that each metallicity production rate is calculated as 

\begin{equation}
   \dot{Z}_{\rm process} = \frac{1}{\rho}\left(\frac{d\rho_Z}{dt}\right)_{\rm process}
\end{equation}

\noindent where $\rho$ is the local gas density and $\left(\frac{d\rho_Z}{dt}\right)_{\rm process}$ is the change in the local density of metals due to the relevant production process. 

This MC method ensures that the tracers used to sample pathlines statistically reproduce the density distribution of the simulation by being probabilistically exchanged between the cells used to discretize the fluid dynamics equations. The probabilistic nature of their exchange between cells prevents the pathlines from being strictly Lagrangian, and introduces some noise in their sampling of the fluid flow.

However, this is preferable to the severe biases introduced with the alternative ``velocity tracer'' method, wherein tracers are evolved by directly integrating their motion by sampling an interpolated velocity field of the simulation. While this algorithm produces smoother pathlines, its bias is so severe that by the end of the  simulation, \textit{all $10^4$} tracers are located at the same position in the simulation -- the center of the simulated galaxy. In contrast the MC tracers sample the full density distribution in the ISM of the galaxy \textit{by design} throughout the entire duration of the simulation. This is consistent with the issues identified in \citet{Genel2013}, and thus MC tracers are the only viable method for sampling pathlines in our simulation.

\begin{figure*}
    \centering
    \includegraphics[width=0.85\textwidth]{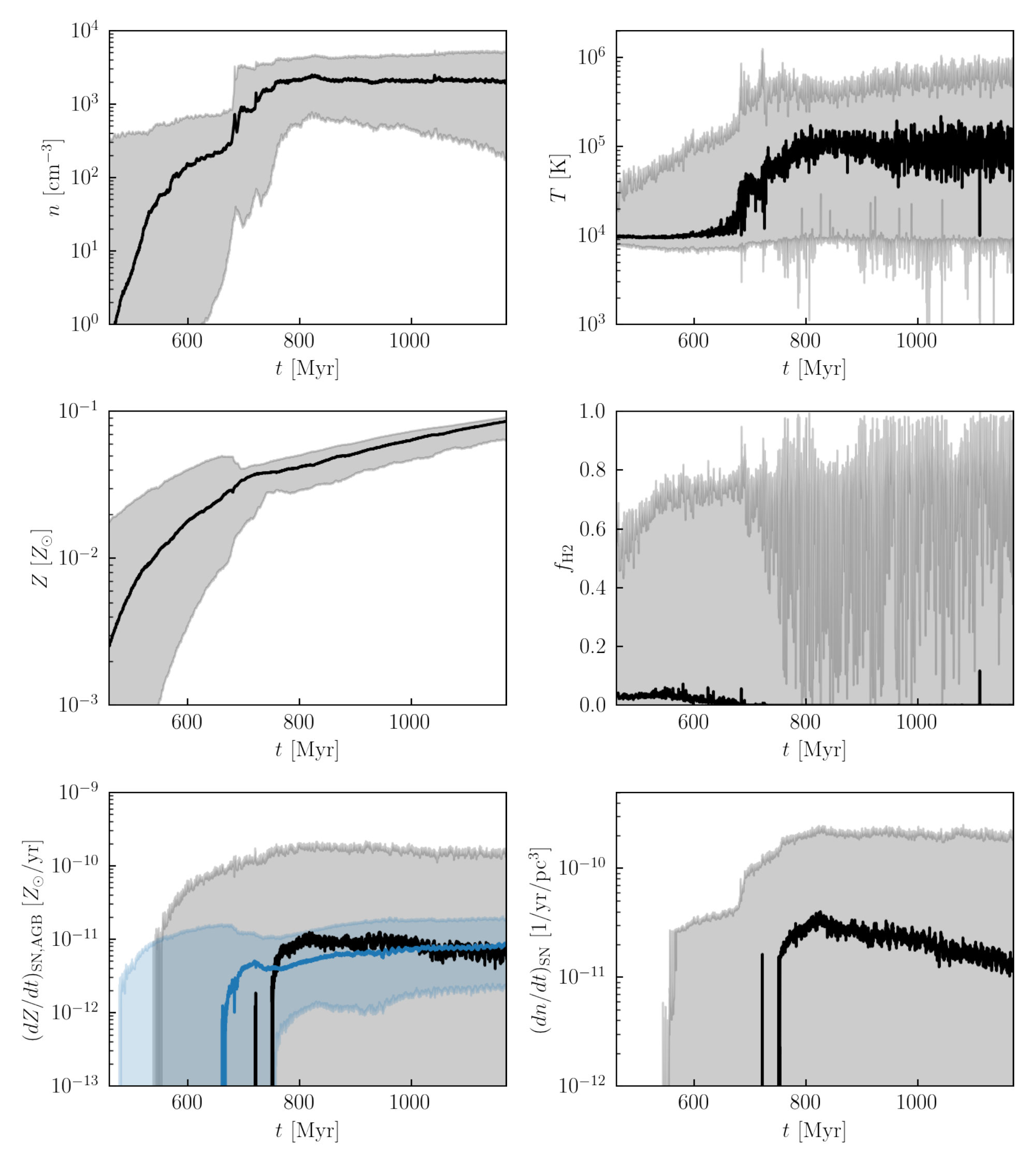}
    \caption{Tracer Histories. Solid black lines indicate median tracer quantities as a function of cosmological time. Shaded regions bound the 16th and 84th percentile tracer quantities. The top right panel additionally shows the molecular-fraction weighted average tracer temperature $\langle T\rangle_{\rm H_2}$ in blue. The bottom left panel shows the metallicity production rates due to AGB stars in blue.}
    \label{fig:tracer_average_histories}
\end{figure*}

Figure~\ref{fig:tracer_average_histories} shows the median value (solid line) and $16$th to 84th percentile range (shaded region) for all tracer properties as a function of time.


\subsection{Dust Model: Physics - Creation and Destruction Processes}

Along each pathline we integrate an ordinary differential equation describing the evolution of the dust-to-gas ratio $D$:

\begin{equation}\label{eq:all_terms}
    \frac{dD}{dt} = \dot{D}_{\rm model} = \dot{D}_{\rm prod} + \dot{D}_{\rm accr} + \dot{D}_{\rm dest}
\end{equation}

\noindent where each term accounts for a different physical process: $ \dot{D}_{\rm prod}$ for production by stellar evolution processes, $\dot{D}_{\rm accr}$ for accretion of gas-phase metals onto dust, and $\dot{D}_{\rm dest}$ for destruction processes that dissociate grains and return metals to the gas phase. We explain how each are calculated in our model in the subsections below. Table \ref{table:params} summarizes all of the free parameters in this model and their default values.

\subsubsection{Dust Production in Evolved Stars and Supernovae}
$\dot{D}_{\rm prod} = \dot{D}_{\rm SN} + \dot{D}_{\rm AGB}$ accounts for the production of dust, which we assume to come from AGB star outflows and SN ejecta. We parameterize each of these as constant fractions $y_D$ of the total metal yields from these processes, so that

\begin{equation}
    \dot{D}_{\rm SN}(\dot{Z}_{\rm SN}|y_{D, \rm SN}) = y_{D, \rm SN}\dot{Z}_{\rm SN}
\end{equation}

\begin{equation}
    \dot{D}_{\rm AGB}(\dot{Z}_{\rm AGB}|y_{D, \rm AGB}) = y_{D, \rm AGB}\dot{Z}_{\rm AGB}
\end{equation}

\noindent where we differentiate between dynamical quantities sampled from the simulation and assumed parameters. 

The complexity of dust grain nucleation and growth in both AGB winds and SN ejecta remnants is still lacking a complete first-principles theoretical description -- see section 4.2 of \citet{HofnerOlofsson2018} for AGB dust formation, \citet{SarangiMatsuuraMicelotta2018} and \citet{MichelottaMatsuuraSarangi2018} for SN dust formation. Observational constraints on the total mass of dust produced by AGB stars exist \citep{Lagadec2008, Srinivasan2009, Matsuura2009, Riebel2012, Nanni2019}, but the dust-to-metal ratio of these outflows is difficult to measure. Large amounts of dust have been observed in SNRs \citep{Dunne2003, Gomez2012, DeLooze2017, Matsuura2011}, but the fraction that survives sputtering after being engulfed by the reverse shock is uncertain. Consequently, neither theoretical nor observational constraints on $y_{D, {\rm AGB/SN}}$ are particularly robust, and we treat these both as free parameters. 

Previous efforts to model dust in galaxy simulations \citep[e.g.][]{McKinnon2016, Li2019} have used the prescription from \citet{Dwek1998} for the production of dust due to AGB and SN, to some success. To obtain an order-of-magnitude estimate for $y_D$, using equations 22-24 of \citet{Dwek1998}, adopting the condensation efficiencies from \citet{Li2019} (based on \citet{FerrarottiGail2006} for AGB, \citet{BianchiSchneider2007} for SN), and taking protosolar elemental abundances \citep{Asplund2009, DraineBook} gives $y_{D,{\rm AGB}} \approx 0.07$ and $y_{D,{\rm SN}} \approx 0.08$. So for default values we take $y_{D,{\rm AGB}} = y_{D,{\rm SN}} = 0.1$, and systematically vary both between $10^{-2}$ and 0.4. We note that, unlike the \citet{Dwek1998} prescription, we do not adjust yields based on the metallicities of stellar populations, since we are not following the abundances of individual elements.

\subsubsection{Gas-Phase Accretion}

$\dot{D}_{\rm accr}$ accounts for dust growth in the ISM due to the accretion of metals onto grains from the gas phase. Kinetic considerations \citep[see][]{Draine1990, Dwek1998,  WeingartnerDraine1999} suggest that this rate has the functional form \citep{Feldmann2015}

\begin{multline}
\label{eq:dotD_acc}
\dot{D}_{\rm accr}(D, Z, n, T, f_{\rm cold} | C_{\rm accr}, \tau_{\rm accr}, f^{\rm dep} ) =\\ \frac{C_{\rm accr}}{\tau_{\rm accr}} f_{\rm cold}\left(\frac{n}{{\rm cm}^{-3}}\right) \left(\frac{T}{{\rm K}}\right)^{1/2} D (f^{\rm dep}Z - D)
\end{multline}

\noindent where $C_{\rm accr}$ is a free parameter to account for uncertainties in the dust clumping on scales unresolved in the simulations, $\tau_{\rm accr}$ is the characteristic timescale, $f_{\rm cold}$ is the fraction of cold gas in which accretion can occur, and $f^{\rm dep}$ is the fraction of metals that can be depleted onto dust grains. Since $\tau_{\rm accr}$ depends upon the average dust grain cross section for each gas phase element that can potentially accrete, and is therefore likely dependent on the detailed composition and geometry of the dust grain population, any theoretical estimate of its value is extremely uncertain. However, \citet{Feldmann2015} showed that the scaling of the galactic dust-to-gas ratio with metallicity depends sensitively on the quantity $t_{\rm dep,H_2}/t_{\rm ISM}$, where $t_{\rm dep,H_2}$ is the molecular depletion timscale and $t_{\rm ISM}$ is their characteristic dust accretion timescale. Using equilibrium and dynamical one-zone models of galactic chemical and dust evolution compared to observations of local galaxies, they constrained $t_{\rm ISM}\approx 4\times 10^4$yr. Their characteristic timescale can be related to ours as 

\begin{equation}
   \tau_{\rm acc} = t_{\rm ISM}\left\langle \left(\frac{n_{\rm cold}}{{\rm cm}^{-3}}\right) \left(\frac{T_{\rm cold}}{\rm K}\right)^{1/2}\right\rangle
\end{equation}

\noindent where $n_{\rm cold}$, $T_{\rm cold}$ are the galaxy-averaged density and temperature of the cool ISM in which grain growth takes place, respectively. These factors are included here since the \citet{Feldmann2015} one-zone modeling did not explicitly account for the density and temperature structure of the ISM in their calculations. Assuming growth takes place in the cold molecular phase lets us estimate $\langle n_{\rm cold}\rangle \approx 10^3{\rm cm}^{-3}$ and $\langle T_{\rm cold}\rangle \approx 50$K \citep{DraineBook}. We therefore adopt as our default value $\tau_{\rm acc} = 3\times10^8$yr with $C_{\rm acc} = 1$. Since this constraint was obtained from galaxy-averaged one-zone models compared only to local-universe observations, it is likely very uncertain, especially in our application to the predictions of high-$z$ galaxies. To explore the effects of this uncertainty, we vary the clumping factor $C_{\rm acc}$ from $0$ to $10$. We adopt $f^{\rm dep} = 0.7$ as in \citet{Feldmann2015} for our default value, and vary from $0.4$ to 1. 

There are two more uncertainties in the calculation of Eq.~\ref{eq:dotD_acc}, both due to the inability of the simulation to resolve the thermodynamic phase structure of the ISM:  the appropriate values for $T$ and $f_{\rm cold}$. The 100 pc minimum cell size of the simulation means that the resolution of the ISM phases will be marginal at best, and the use of a delayed cooling feedback model risks over-predicting the temperature of the high-density gas. As we show in \ref{subsubsec:ISM_phase_structure}, the phase structure of the ISM is likely not well resolved in this simulation. Therefore, naively using the simulation temperature for Eq.~\ref{eq:dotD_acc} would grossly over-predict the ISM growth rate. Consequently, by default we assume $T = T_{\rm cold} = 50$K for Eq.~\ref{eq:dotD_acc} only, but we also present the predictions of the model when the full simulation temperature is used. Since this unresolved cold phase represents a subset of the total gas in each cell, we must also multiply by the mass fraction of cold gas $f_{\rm cold}$. We can estimate the value of this additional free parameter by equating it to either the neutral fraction $f_{\rm HI}$ which is calculated self-consistently from the radiation field, density, and temperature of the ISM in the simulations, or the molecular fraction $f_{\rm H2}$ modeled with the \citet{GnedinDraine2014} fitting functions. We use $f_{\rm cold} = f_{\rm H2}$ from the simulation by default, since this is the same cold fraction assumed by the star formation prescription of the simulation. However, we also explore the predictions for $f_{\rm cold} = f_{\rm HI}$.

\begin{figure*}
    \centering
    \includegraphics[width=\textwidth]{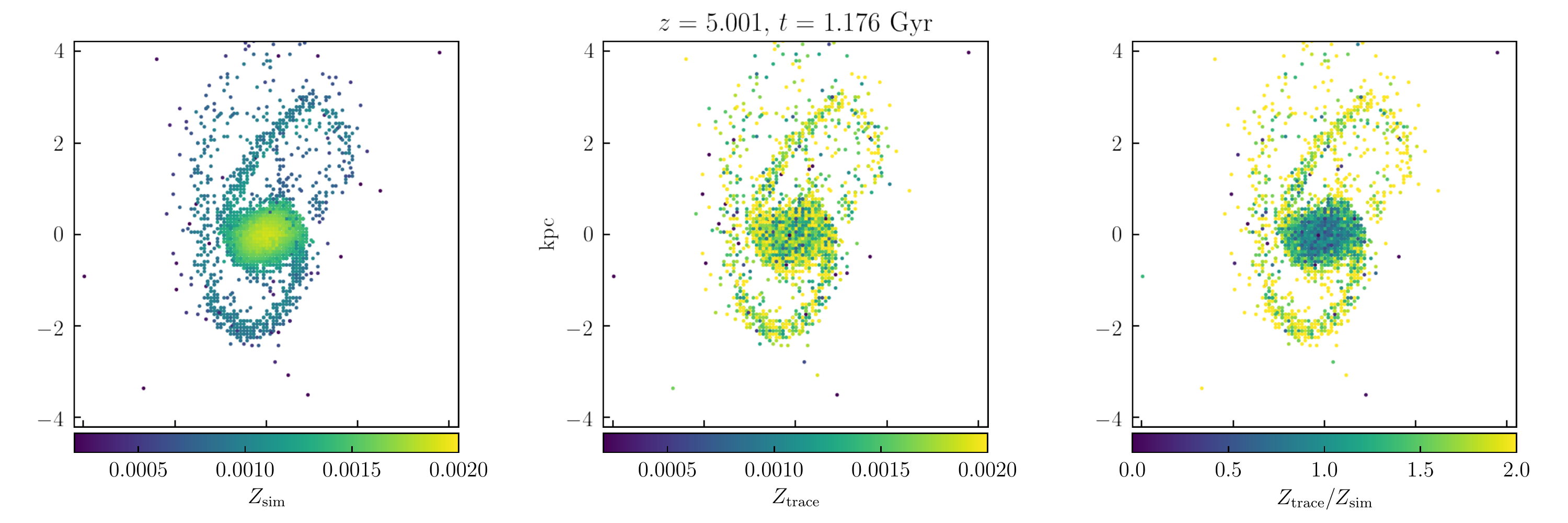}
    \caption{Maps of the simulation metallicity (left panel), tracer metallicity (defined by Eq.~\ref{eq:tracer_metallicity}, middle panel), and their ratio (right panel) at the final snapshot of our simulation are shown. Values for tracers in the same cell are averaged. The statistical error in the tracer paths due to their probabilistic nature results in an approximately uniform tracer metallicity across the galactic disk, while the simulation metallicity displays a clear gradient.}
    \label{fig:Ztrace_vs_Zsim_map}
\end{figure*}

\subsubsection{Destruction Processes}
\label{subsubsec:dest}
$\dot{D}_{\rm dest}$ is the rate at which dust is destroyed by energetic processes in the ISM. The most important of these is likely thermal sputtering in high temperature gas -- the erosion of dust grains due to thermal collisions with high-temperature gas particles. The timescale for the erosion of graphite, silicate, and iron grains of size $a$ due to thermal sputtering is given by \citep[Eq. 25.14]{DraineBook} 

\begin{equation}
\frac{da}{dt} = -10^{-6}{\rm \mu m}\;{\rm yr}^{-1}\left[1 + \left(\frac{T}{10^6{\rm K}}\right)^{-3}\right]\left(\frac{n_{\rm H}}{\rm cm^{-3}}\right)
\end{equation}

\noindent which we can convert to change in dust-to-gas ratio (assuming a single size for all grains) as 

\begin{equation}
\frac{dD}{dt} = \frac{3}{a}D\frac{da}{dt}
\end{equation}

\noindent giving

\begin{multline}\label{eq:sputtering}
\dot{D}_{\rm sput}(D, n, T | C_{\rm sput}, \tau_{\rm sput}, \omega) =\\ -\frac{C_{\rm sput}}{\tau_{\rm sput}}\frac{1}{1 + (T/10^6{\rm K})^{-\omega}}\left(\frac{n}{{\rm cm}^{-3}}\right) D
\end{multline}

\noindent where $C_{\rm sput}$ is a free parameter of default value unity we systematically vary to account for uncertainties, $\tau_{\rm sput} \approx 3\times10^5{\rm yr}(a/{\rm \mu m})$ is the characteristic timescale for dust destruction due to sputtering in high temperature gas, and $\omega \approx 3$ parameterizes the temperature dependence of the sputtering rate for $T \ll 10^6$K. Adopting $a= 0.1\mu$m \citep{DraineBook}, we take $\tau_{\rm sput} = 3\times 10^4$yr and $\omega = 3$ as our default values.

If the temperature and density distribution of the ISM were fully resolved in our simulations, thermal sputtering alone would provide a good estimate of the grain destruction rate. However, since the low resolution and delayed cooling feedback in the simulation likely over-predict the temperature of the densest ISM gas, Eq~\ref{eq:sputtering} with quantities from the simulation likely grossly over-predicts the destruction rate in the ISM. 

To correct for this, an alternative method of calculating the destruction rate is tied directly to the local rate of supernovae, which are assumed to be responsible for all of the $\sim 10^6$K gas in the ISM where sputtering is most efficient. Let $M_{\rm SNR}$ be the amount of ISM mass swept up by an individual SNR, and $C_{\rm dest}$ be the average efficiency with which grains are destroyed in each SNR. Then the rate of change in dust-to-gas ratio is given by 

\begin{equation}\label{eq:dotD_SNR}
    \dot{D}_{\rm SNR}(D, n, \dot{n}_{\rm SN}|C_{\rm dest} M_{\rm SNR}) = C_{\rm dest} M_{\rm SNR}D\frac{\dot{n}_{\rm SN}}{\rho}
\end{equation}

\noindent where $\dot{n}_{\rm SN}$ is the local volumetric rate of supernova explosions and $\rho$ is the gas density. \citet{McKee1989} estimate $M_{\rm SNR} \approx 1000M_{\odot}$, which we adopt for our fiducial value. $C_{\rm dest}$ is a free parameter with a fiducial value of unity that we vary to explore the impact of uncertainties in this rate due to both the uncertain destruction efficiency and $M_{\rm SNR}$. 

Since $ \dot{D}_{\rm SNR}$ is not dependent on the local gas temperature, which is likely inaccurate in the simulation due to limited resolution, we adopt $\dot{D}_{\rm des} = \dot{D}_{\rm SNR}$ by default, but also explore predictions using $\dot{D}_{\rm des} = \dot{D}_{\rm sput}$.

\begin{deluxetable*}{|c|c|c|}
\setlength{\tabcolsep}{1.5pt}
\startdata
\tablehead{Model Parameters & Description &{\bf Default} (Values Explored ) }
%
Production, $\dot{D}_{\rm prod}$ & &\\
\hline
$y_{D, {\rm SN}}$ & Yield (as a fraction of $Z$) from Supernovae & {\bf 0.1}(0 -- 0.4)\\
$y_{D, {\rm AGB}}$ & Yield (as a fraction of $Z$) from AGB & {\bf 0.1}($10^{-2}$ -- 0.4)  \\
\hline
Gas-Phase Accretion, $\dot{D}_{\rm accr}$ & &\\
\hline 
$C_{\rm accr}$ & Clumping/Uncertainty Factor &  {\bf 1}($0$ -- $10$)\\
$\tau_{\rm accr}$ & Growth Timescale & {\bf $3\times 10^{8}$yr}\\
$f^{\rm dep}$ & ``Depletable" Medal Fraction & {\bf 0.7}(0.4 -- 1)\\
$f_{\rm cold}$ & Sub-grid cold gas fraction &  {\bf $f_{\rm H_2}$} ($f_{\rm HI}$, 1)\\
$T_{\rm cold}$ & Assumed Temperature of Cold ISM & {\bf 50 K} (Direct from Simulation).\\
\hline
\hline
Destruction: SNR, $\dot{D}_{\rm dest} = \dot{D}_{\rm SNR}$ & & \\
\hline
$C_{\rm dest}$ & Destruction Efficiency/Uncertainty Factor & {\bf 1} ($0$ -- $10$) \\
$M_{\rm SNR}$ & ISM Mass Swept-Up by Each Supernova Remnant & {\bf $1000 M_{\odot}$}\\
\hline
Destruction: Thermal Sputtering, $\dot{D}_{\rm dest} = \dot{D}_{\rm sput}$& & \\
\hline
$C_{\rm sput}$ & Uncertainty Factor & {\bf 0} (0-10)\\
$\tau_{\rm sput}$ & Destruction Timescale at $T = 10^6{\rm K}$ &  {\bf $3\times 10^{-2}$ Myr}\\
$\omega$ & Temperature Dependence Coefficient & {\bf 3} \\
\enddata
\caption{Model Parameters}
\label{table:params}
\end{deluxetable*}


\subsection{Dust Model: Numerical Integration}\label{sec:numerical_methods}

\begin{figure}
    \centering
    \includegraphics[width=\linewidth]{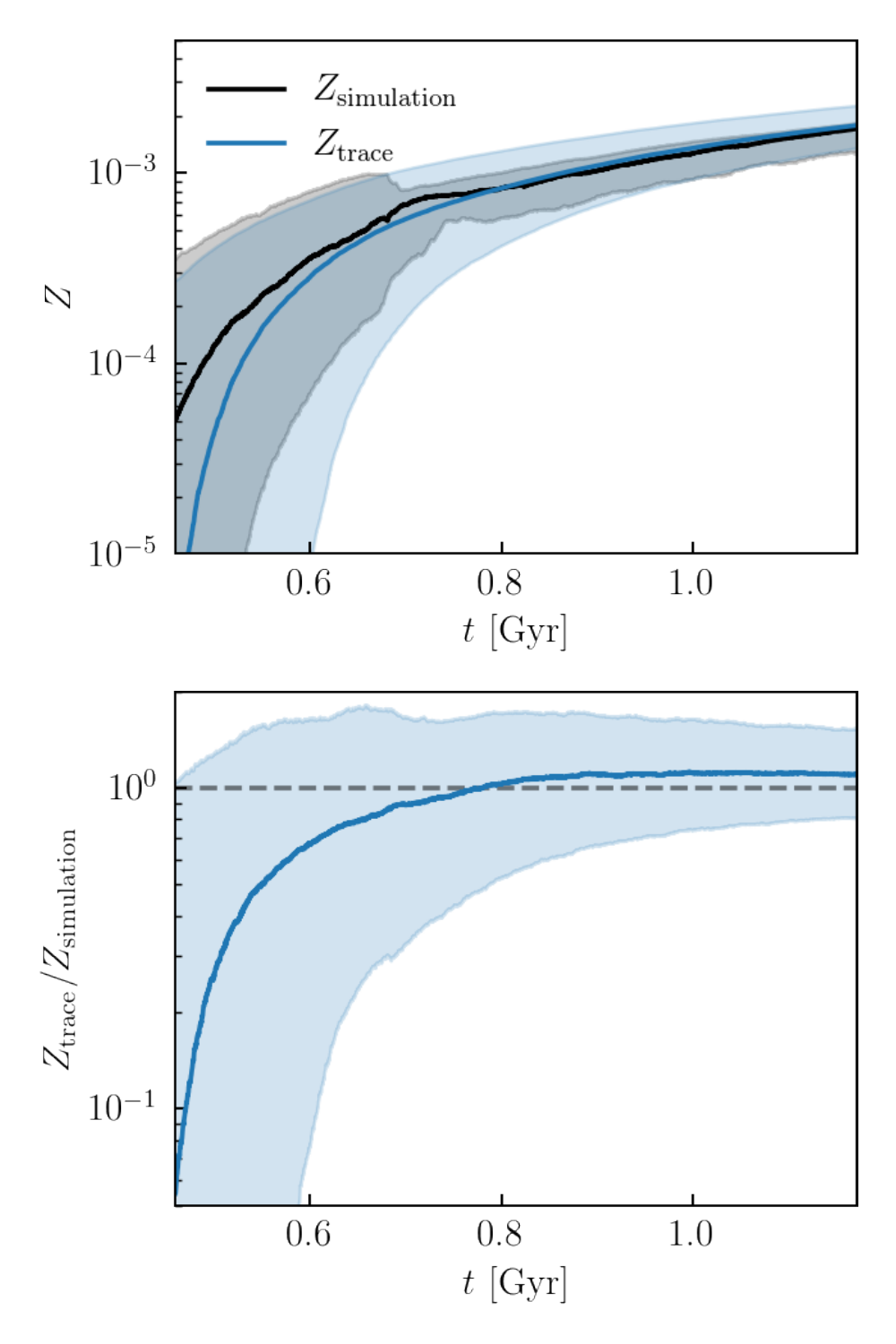}
    \caption{The top panel shows the evolution of both simulation (black) and tracer (blue) metallicity as a function of time in the simulation, while the bottom shows their ratio. The solid lines in the top pannel show the median tracer value at each snapshot, while the shaded regions enclose the 16th and 84th percentiles. The bottom shows the same for their ratio. By $\sim 800${\rm Myr}, their ratio is close to unity, indicating that the tracer metallicity reflects the simulation metallicity in an average sense. Before that time, however, the tracer metallicity is biased low. We discuss the correction of this bias in Section~\ref{sec:numerical_methods}.}
    \label{fig:Ztrace_vs_Zsim}
\end{figure}

For each tracer $i$ and each snapshot $j$ in the tracer data, we have the following quantities from the simulation: $n_{ij}$, $T_{ij}$, $f_{{\rm HI},ij}$, $f_{{\rm H_2},ij}$, $\dot{Z}_{{\rm SN},ij}$, $\dot{Z}_{{\rm AGB},ij}$, $\dot{n}_{{\rm SN},ij}$. For each tracer, we linearly interpolate each quantity $X$ as a function of cosmological time to define a continuous function $X_i(t)$. This allows us to define a pair of coupled continuous ordinary differential equations for the metallicity and dust-to-gas ratio of each tracer:

\begin{equation}\label{eq:tracer_metallicity}
    \frac{dZ_i}{dt} = \dot{Z}_{{\rm SN},i}(t) + \dot{Z}_{{\rm AGB},i}(t)
\end{equation}

\begin{multline}\label{eq:dust_ode_0}
    \frac{dD_i}{dt} = \dot{D}_{\rm model}[n_i(t), T_i(t), f_{{\rm HI}, i}(t), f_{{\rm H_2}, i}(t), Z_i(t),\\ \dot{Z}_{{\rm SN},i}(t), \dot{Z}_{{\rm AGB},i}(t), \dot{n}_{{\rm SN},i}(t)].
\end{multline}

\noindent We use the scipy \citep{2020SciPy-NMeth} module \texttt{solve\textunderscore ivp} to perform this numerical integration. We have tested our solver against subsets of the model with analytically expressible solutions, all of which are recovered to fractional accuracy of $\lesssim 10^{-7}$ with the adopted solver parameters.

Note that $Z_i$ is {\it not} generally identical to the metallicity in the simulation sampled by the tracer, call it $Z_{{\rm sim},i}$, because of the imperfectly Lagrangian nature of the Monte Carlo tracers (see Section~\ref{subsec:methods_tracers}). Indeed, Figure~\ref{fig:Ztrace_vs_Zsim_map} compares the spatial distribution of these two quantities in the last snapshot of our simulation, at $z=5$. While the simulation predicts a smooth radial gradient in the metallicity of the galaxy disk, the integrated production is both noisy and uniform across the disk due to the probabilistic nature of the tracers. 

A more quantitative comparison is shown in Figure~\ref{fig:Ztrace_vs_Zsim}, where the median, 16th and 84th percentiles of these metallicities for all tracers at each time are compared. By the end of the simulation, their ratio is close to unity, indicating the tracers capture the global metal production history of the ISM in the galaxy, but due to sampling issues are biased low at times earlier than $\sim 600$ Myr. However, this effect can be corrected for: since the simulation metallicity {\it is} Lagrangian (because it is evolved with an advection solver in the simulation code), we can re-scale the dust-to-gas ratio of each tracer by the ratio of these metallicities. So, for any observable quantity that requires a dust fraction $D_{\rm obs}$,  this is calculated from the $D$ predicted by our model as

\begin{equation}\label{eq:D_obs}
D_{\rm obs} = \left(\frac{D}{Z}\right)_{\rm tracer}Z_{\rm sim}
\end{equation}

\subsection{Correction of Star Formation Rates}\label{subsec:SFR_corr}

\citet{Zhu2020} showed that the CROC simulations fail to reproduce the observed stellar mass-halo mass relationship. The stellar masses predicted by the simulations, and therefore star formation rates, are too low at a given halo mass compared to the observations. They provide an empirical formula that re-scales the simulated stellar masses to be consistent with observations:

\begin{multline}\label{eq:SFR_corr}
    \tilde{M}_{\star} = M_{\star}\left[1 + A_z(t)\log_{10}\left(1 + \frac{M_{\star}}{3\times10^8M_{\odot}}\right)\right] \\ \equiv M_{\star}f_{\rm corr}(t)
\end{multline}

\noindent where $A_z = 6, 10, 30$ at $z = 5, 6, 7$. We test the sensitivity of our dust model prediction to this inaccuracy in the simulation by appropriately re-scaling the metal production rates ($\dot{Z}_{\rm SN}$, $\dot{Z}_{\rm AGB}$) and supernova rates $\dot{n}_{\rm SN}$ by this same factor: $\dot{\tilde{Z}}_{\rm SN} = f_{\rm corr} \dot{Z}_{\rm SN}$, $\dot{\tilde{Z}}_{\rm AGB} = f_{\rm corr} \dot{Z}_{\rm AGB}$, and $\dot{\tilde{n}}_{\rm SN} = f_{\rm corr} \dot{n}_{\rm SN}$. Note that this requires re-scaling the simulation metallicity for the correction in Eq.~\ref{eq:D_obs}. We interpolate $f_{\rm corr}$ to all simulation redshifts by fitting a second-order polynomial to $A_z$. Since the instantaneous metal production and supernova rates depend on the entire current stellar population, they depend on the full prior star formation history. Our correction is therefore a crude approximation. Moreover, substantially increased star formation would also remove additional gas from the galaxy ISM, thereby likely changing the density distribution in a complicated way we do not account for. Consequently, this correction should not be taken as a precise prediction, but rather as providing an estimate of the uncertainty in our predictions due to the inaccuracies in the simulation. 

\subsection{Calculation of Observable Quantities: $M_D$, $\beta_{\rm UV}$, IRX, $\tau_{1500}$}
To compare our predictions to observational data, we calculate quantities of our simulated galaxy that depend on dust content and can in principle be measured observationally. These predictions require as input the spatial distribution of the dust mass, which we obtain as follows. For each simulation snapshot we extract a fixed resolution data cube centered on the simulated galaxy with dimensions of $L = 0.2R_{\rm vir}$ on each side. The data cube has a resolution equivalent to the highest refinement level of the adaptive mesh, since this is the resolution of the ISM gas that contains almost all the dust. Each tracer particle is then identified with its corresponding cell. In order to adequately sample the gas distribution in the simulation, the simulation was run with sufficiently many tracers that most cells in the simulated galaxy ISM have multiple tracers. For each cell we calculate the dust-to-gas ratio as the average of all tracers in the same cell. We can then multiply this quantity by the cell metallicity and the cell mass to obtain the dust mass per cell.

The total dust mass is simply the sum of individual cell dust masses. While this is the most straightforward measure of the galactic dust content, it is difficult to measure the dust mass of high-redshift galaxies, as this relies on assumptions of the grain size distribution and dust temperature. Consequently, we also forward-model several more directly observable quantities: the attenuated ultravoilet luminosity $L_{\rm UV}$, the effective optical depth due to dust extinction in the ultraviolet (at $\lambda = 1500\AA$) $\tau_{1500}$, the logarithmic slope of the galaxy spectrum in the ultraviolet $\beta_{\rm UV}$, and the infrared luminosity due to dust thermal radiation $L_{\rm IR}$ and corresponding infrared excess IRX$\equiv L_{\rm IR}/L_{\rm UV}$. 

The ultraviolet radiation from galaxies comes from young massive stars, but can be significantly attenuated by interstellar dust. This attenuation is wavelength-dependent such that the shape of the spectrum in this regime is sensitive to the dust column density. $\beta_{\rm UV}$ is calculated by least-squares fitting a power law of the form $f_{\lambda} \propto \lambda^\beta$ to the galaxy spectrum in the rest-frame wavelengths $1268-2580\AA$, using only the 10 wavelength windows listed in Table 2 of \citet{Calzetti1994} to avoid contamination from absorption lines. The galaxy spectrum is the combined spectra of each star particle, each of which represents a single-age, uniform-metallicity stellar population attenuated by a dust column density dependent on the observational viewing angle. The unattenuated spectrum of each star particle is calculated using the Flexible Stellar Population Synthesis code \citep{ConroyGunn2010}. To sample the distribution of viewing angles, we calculate optical depths projected along the 3 coordinate axes in both directions (6 total), which should be random with respect to the galaxy orientation. Optical depths to each star particle as function of wavelength are estimated as

\begin{equation}
    \tau_i(\lambda) = \kappa_D(\lambda)\int_i\rho_D dl
\end{equation}

\noindent where $\kappa_D(\lambda)$ is the opacity (or mass absorption coefficient) {\it per unit dust mass}, $\rho_D$ is the dust density, and $\int_idl$ is the intergal along a line-of-sight to a given star $i$. For  $\kappa_D(\lambda)$ we use the ``SMC bar'' model of  \citep{WeingartnerDraine2001}.\footnote{\url{https://www.astro.princeton.edu/~draine/dust/dust.html}, converted from a total mass to dust mass absorption coefficient with an assumed dust-to-gas mass ratio of $M_{\rm dust}/M_{\rm gas} = (M_{\rm dust}/M_{\rm H})/(M_{\rm gas}/M_{\rm H}) = 0.00206/1.36 = 0.00151$ from Table 3 of \citet{Draine2007}.} 

Dust grains are heated to temperatures $T\sim10-100$K by the interstellar radiation field, casing them to emit thermal radiation in the infrared. Assuming the ISM is optically thin to this radiation, the specific luminosity is given by

\begin{equation}
    L_{\lambda} = 4\pi M_D \kappa_D(\lambda) B_{\lambda}(T)
\end{equation}

\noindent  where $\kappa_D(\lambda)$ is the dust opacity, $B_{\lambda}(T)$ is the specific intensity of black-body radiation. Since this quantity is dependent on the dust temperature, a fully self-consistent calculation would require solving for the dust temperature given the joint spatial distributions of dust and interstellar radiation. Since our simulations likely do not resolve this structure, a completely self-consistent calculation of this quantity is impossible. Consequently, we present results for several temperatures that span a plausible range: 20, 40, 60K. As in \citet{Bouwens2020}, we define the total infrared luminosity as $L_{\rm IR} = \int_{8\mu m}^{1000\mu m}L_{\lambda}d\lambda$, and estimate the UV luminosity as $L_{\rm UV} = L_{1500\AA}$, to calculate the infrared excess as IRX$\equiv L_{\rm IR}/L_{\rm UV}$.

\section{Results}

\subsection{Predictions of Dust Model: Dependence on Dust Model Parameters}

\begin{figure}
    \centering
    \includegraphics[width=\linewidth]{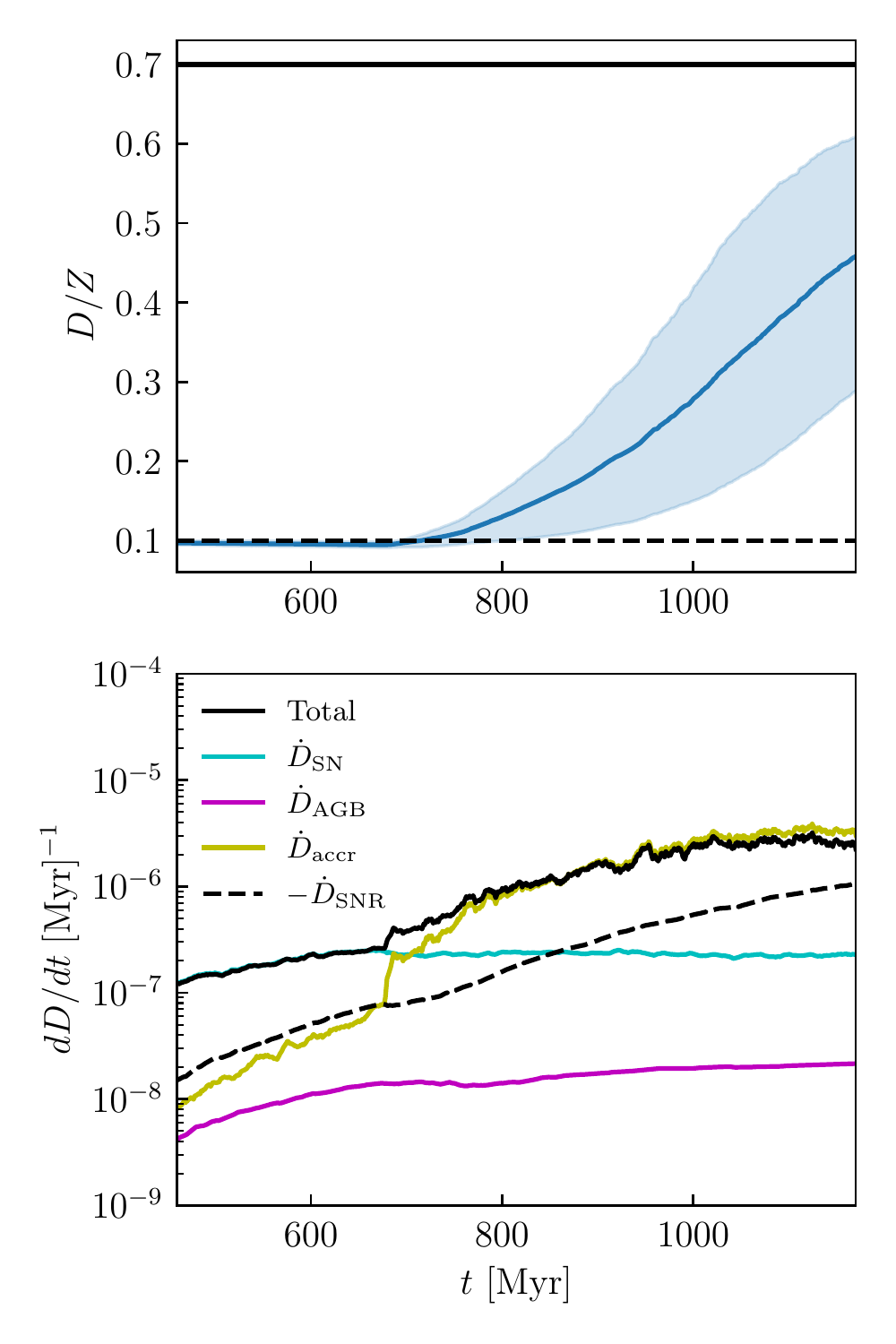}
    \caption{Dust-to-metal ratios (top panel) and the contributions of each process to dust growth with default parameters (bottom panel). Solid lines show mean values for all tracers, and the shaded region on the top panel bounds the 16th and 84th percentiles at each timestep. On the top panel, the solid black line shows $f^{\rm dep}$ and the dashed black line shows $y_{D,{\rm SN}} = y_{D,{\rm AGB}}$. The rapid variation in simulated ISM quantities sampled by the tracers (as shown in Figure~\ref{fig:tracer_average_histories}) result in rapid variation of of these rates, so for readability we have smoothed the lines on the bottom panel with a boxcar average of width 10 Myr.}
    \label{fig:DtoZ_model_summary}
\end{figure}

Fig.~\ref{fig:DtoZ_model_summary} shows the resulting dust-to-metal ratio and rates of individual dust creation and destruction physical processes for our default parameter values. We see that the dust production is initially dominated by supernovae ejecta, in which case the dust-to-metal ratio is set by the assumed supernova yield $y_{D,\rm SN}$. Note that the destruction rate due to supernova remnants is much lower than the production rate, so that $D/Z$ never falls significantly below $y_{D,{\rm SN}}$, indicated with the dashed black line in the top panel. 

At approximately 700 Myr, accretion increases rapidly and dominates the dust evolution thereafter. Consequently, as metals in the ISM accrete onto dust grains, the dust-to-metal ratio now increases from $\sim y_{D,\rm SN}$, reaching a value of approximately $0.5$ by the end of the simulation at $z=5$, $t=1.18$Gyr. Note that this is less than the maximum of $D/Z = f^{\rm dep} = 0.7$. Despite increasing, the rate of supernova remnant destruction always remains subdominant to ISM accretion. The contribution of dust from AGB stars is always negligible because they produce heavy metals at a rate approximately two orders-of-magnitude less than supernovae. 

Below we explore how these predictions change with model parameter values.

\subsubsection{Production: $y_{\rm D, SN}$ and  $y_{\rm D, AGB}$}

\begin{figure}
    \centering
    \includegraphics[width=\linewidth]{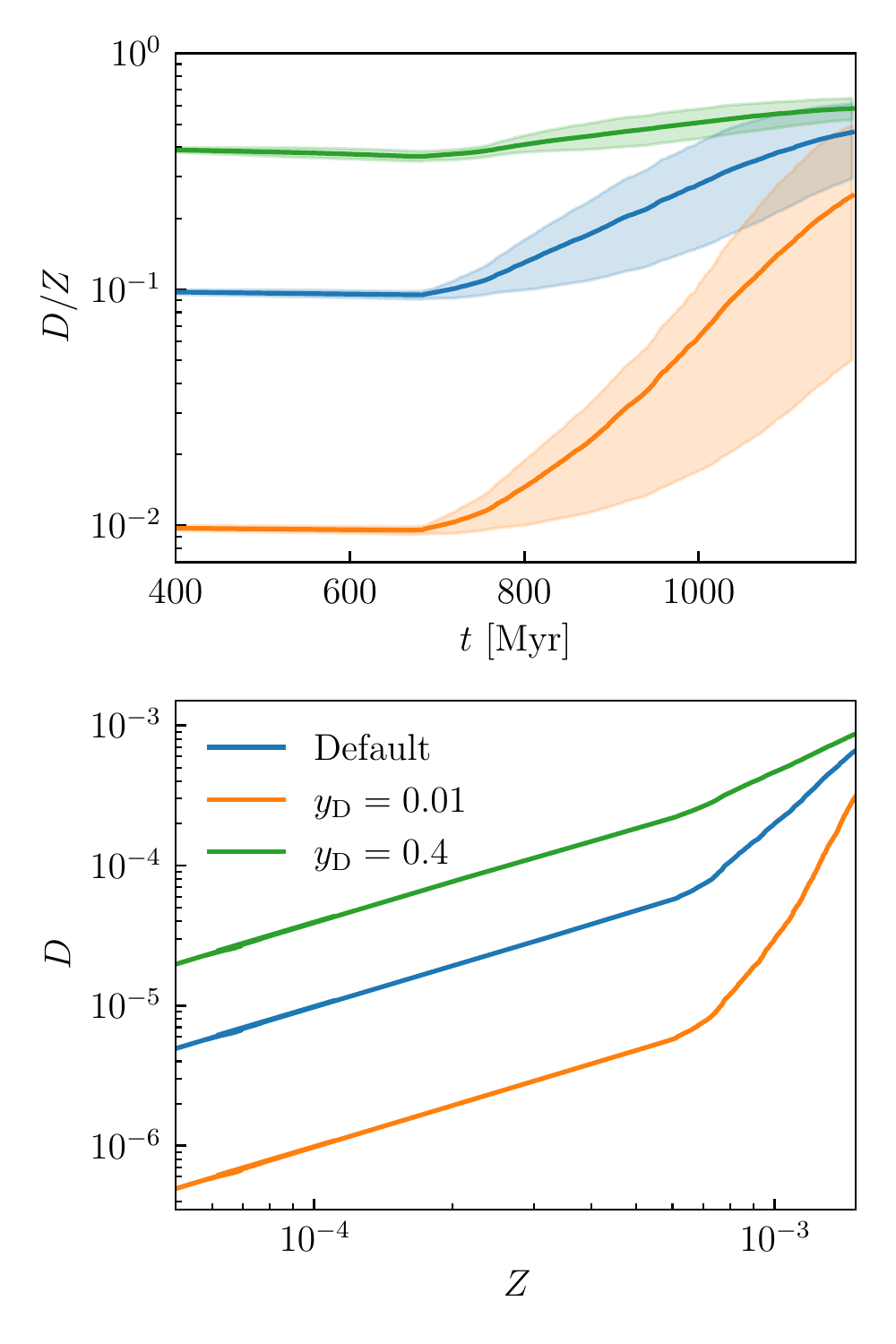}
    \caption{Effect of varying production yield $y_{D,{\rm SN}} = y_{D,{\rm AGB}} = y_D$ on dust evolution. Solid lines show mean values for all tracers, shaded regions encompass tracer 16th and 84th percentiles.}
    \label{fig:DtoZ_DvsZ_yD}
\end{figure}

\begin{figure}
    \centering
    \includegraphics[width=\linewidth]{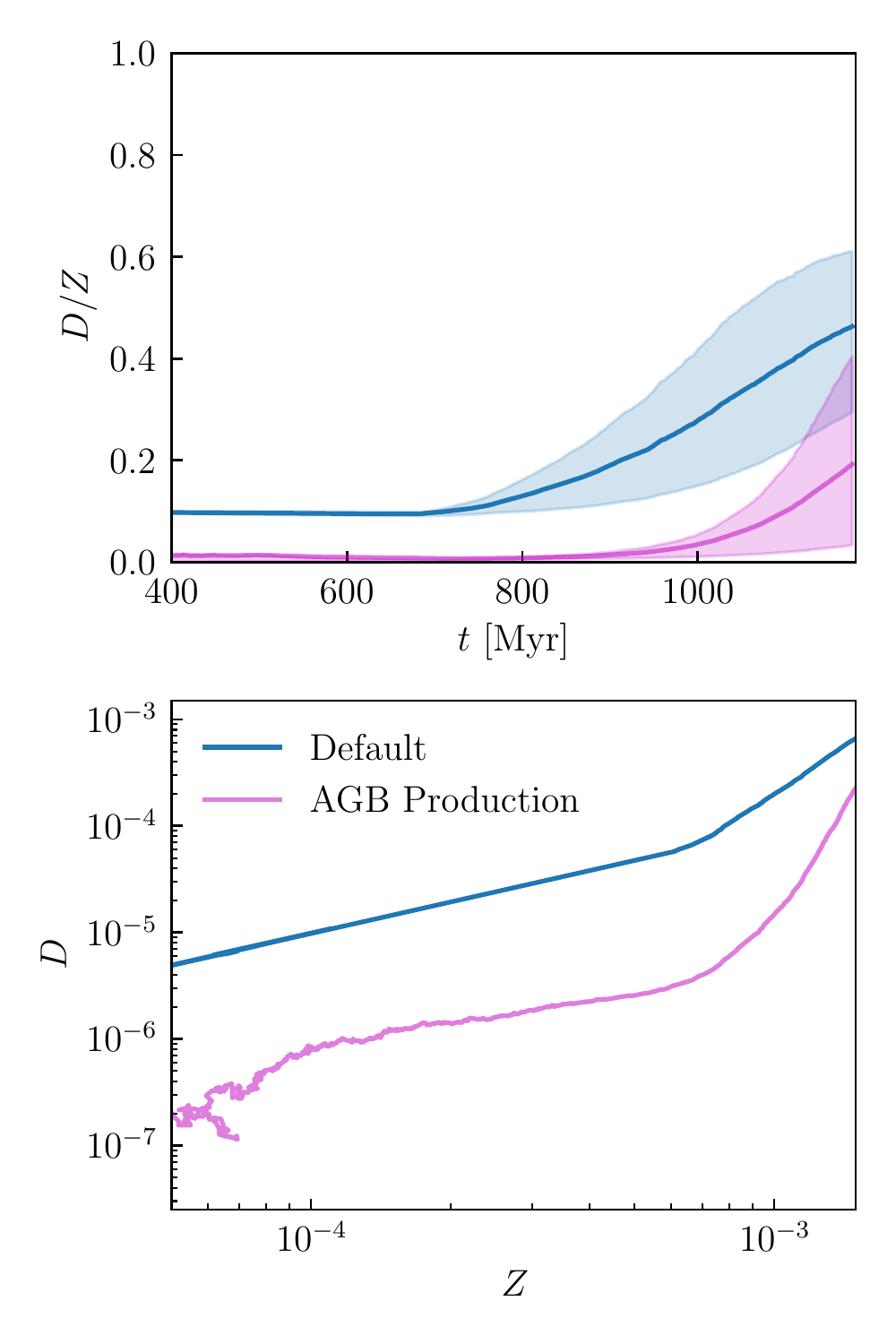}
    \caption{Only AGB dust production. The default value $y_{\rm D, AGB} = 0.1$ is used in both cases, while the magenta lines show the predictions of the model wen SN production is turned off completely $y_{\rm D, SN} = 0$. Solid lines and shaded regions are defined in the same way as the previous figure.}
    \label{fig:DtoZ_DvsZ_no_SN}
\end{figure}

Fig.~\ref{fig:DtoZ_DvsZ_yD} explores the effect of varying the production yield $y_D$ on $D/Z$ and the evolution of $D$ as a function of $Z$. Note that we use this to indicate the value for both $y_{\rm D, SN}$ and  $y_{\rm D, AGB}$, which we keep the same in this figure. As shown in Fig.~\ref{fig:DtoZ_model_summary}, AGB wind production is largely negligible at these redshifts because it produces far fewer metals, so these results are unchanged for any physically reasonable variations in $y_{\rm D, SN}/y_{\rm D, AGB}$ around 1. Values of $y_D = 0.01$ and $y_D = 0.4$ are compared with the default $y_D = 0.1$. In each case the value of $y_D$ entirely determines the dust-to-metal ratio prior to the onset of rapid accretion from the ISM. By setting the intitial condition for this rapid accretion phase, $y_D$ is positively correlated with the final $D/Z$ ratio. However, note that the rapid accretion phase begins at exactly the same time, and therefore same metallicity, in each model. This is consistent with the finding of \citep{Feldmann2015} that accretion becomes efficient at a critical metallicity which is independent of the production yield. Finally, note that because the supernova remnant destruction rate is $\propto D$ (eq.~\ref{eq:dotD_SNR}), destruction (with default parameters) remains unimportant even when the production yield is very small, and the $D/Z$ ratio never goes significantly below $y_D$. 

Fig.~\ref{fig:DtoZ_DvsZ_no_SN} shows the predicted dust content in the absence of dust production by supernovae, i.e. dust is assumed to be produced only from AGB star winds. While young local supernova remnants are observed to efficiently form dust, it remains an open question how much of this dust survives the reverse shock, motivating this parameter choice as an extreme scenario in which supernovae do not produce any dust grains. This figure shows that even with reasonable assumed AGB wind dust yield (10\%) and ISM growth timescale, AGB production {\it alone} produces enough dust to allow efficient accretion of gas phase metals at late times in our simulations, resulting in $D/Z \sim 0.2$ by $z=5$.  

\subsubsection{Gas-Phase Accretion: $C_{\rm accr}$, $T_{\rm cold}$, and $f^{\rm dep}$}

\begin{figure}
    \centering
    \includegraphics[width=\linewidth]{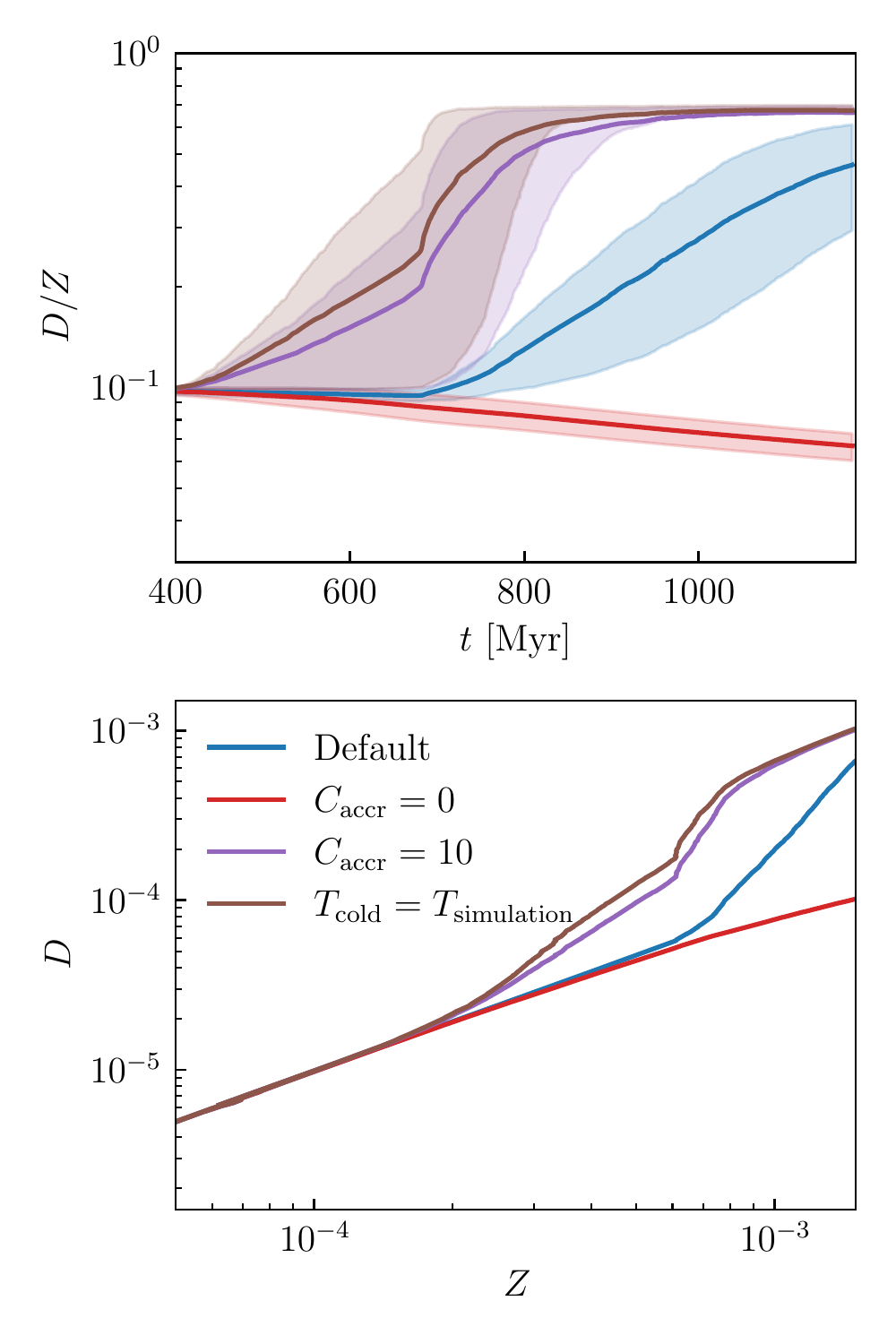}
    \caption{Effect of varying ISM accretion rate uncertainty/clumping factor $C_{\rm accr}$ and the cold phase gas temperature $T_{\rm cold}$ (which are degenerate, as explained in the text).}
    \label{fig:DtoZ_DvsZ_c_accr}
\end{figure}

\begin{figure}
    \centering
    \includegraphics[width=\linewidth]{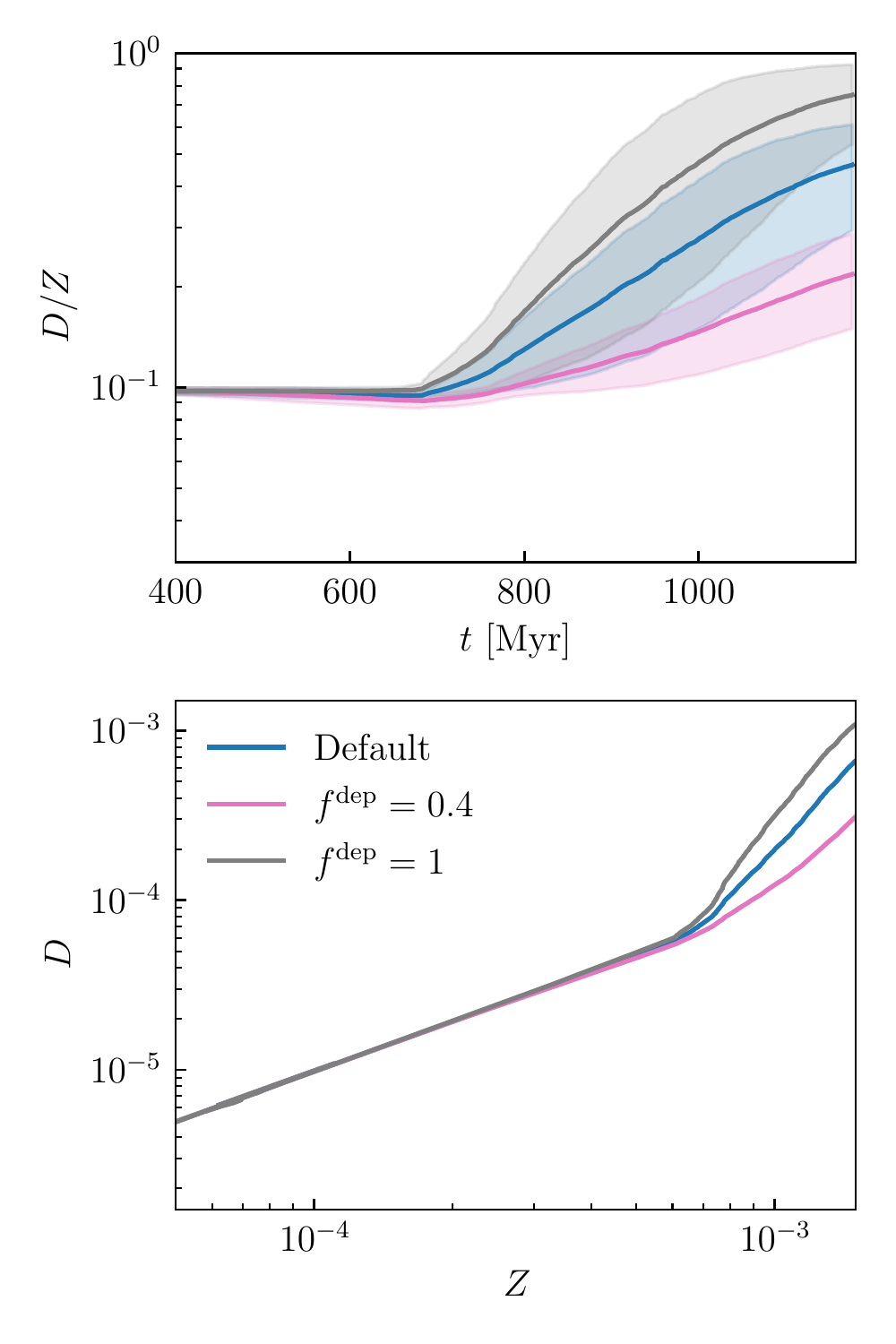}
    \caption{Effect of varying maximum ISM metal depletion fraction $f^{\rm dep}$.}
    \label{fig:DtoZ_DvsZ_fdep_accr}
\end{figure}

Figures \ref{fig:DtoZ_DvsZ_c_accr} and \ref{fig:DtoZ_DvsZ_fdep_accr} explore the effect of changes to gas-phase accretion parameters. Since $\dot{D}_{\rm accr} \propto C_{\rm accr} T_{\rm cold}^{1/2}$ (eq.~\ref{eq:dotD_acc}), we show various choices for both in Fig.~\ref{fig:DtoZ_DvsZ_c_accr}.  Fig.~\ref{fig:DtoZ_DvsZ_fdep_accr} explores variations in the maximum metal depletion fraction $f^{\rm dep}$. 

Fig.~\ref{fig:DtoZ_DvsZ_c_accr} shows three modifications to the accretion prescription in the default model: accretion turned off entirely $C_{\rm accr} = 0$, accretion enhanced by an order-of-magnitude $C_{\rm accr} = 10$, and setting the temperature of the ``cold" phase in which accretion takes place to the local gas temperature of the simulation $T_{\rm cold} = T_{\rm simulation}$ (as opposed to assuming the constant value of $T_{\rm cold} = 50$K as is our default). In the complete absence of ISM accretion, destruction due to supernova remnants is able to reduce the dust-to-metal ratio by $\sim30\%$ compared to the production yield. Conversely, an order-of-magnitude enhancement in the accretion rate causes the dust mass to grow rapidly at earlier times and lower metallicities, saturating at $D/Z \approx f^{\rm dep}$ well before the end of the simulation at $z=5$. 

We see that changing $T_{\rm cold}$ from a constant value of $50$K to the gas temperature in the simulation sampled by each tracer has a similar but even slightly greater effect. Note that since we adopt $f_{\rm cold} = f_{\rm H_2}$, this is effectively the average temperature of the molecular phase predicted by the simulation. As shown in Fig.~\ref{fig:tracer_average_histories}, the molecular-fraction-weighted average temperatures are $\gtrsim 5000$, and the accretion rate depends on the square-root of the ambient gas temperature, so $\sqrt{\langle T_{\rm H_2}\rangle / 50{\rm K}} \gtrsim 10$ is the expected enhancement. In this way the assumed average temperature of the cold phase is degenerate with the assumed timescale for grain growth due to gas-phase accretion, as long as $T_{\rm cold}$ does not exhibit any broad, significant trends on timescales of gigayears, as is the case even when we take the simulation value.

Fig.~\ref{fig:DtoZ_DvsZ_fdep_accr} shows the effect of changing the ``depletable'' metal fraction $f^{\rm dep}$ -- the maximum mass fraction of metals that can be depleted onto dust grains. We adopt a default value of $0.7$ and compare to values of $0.4$ and $1$ which represent extremes allowed by observations -- physically $f^{\rm dep} \le 1$ because the dust mass cannot exceed the metal mass, and $f^{\rm dep} \gtrsim 0.4$ because this is the observed dust-to-metal ratio in the ISM of the Milky Way \citep{Draine2007}. Since this value sets the saturation of $D/Z$ when accretion is dominant, higher $f^{\rm dep}$ corresponds to a higher final $D/Z$ at late times. Therefore the shape of the transition from production dominated $D/Z\sim y_D$ to accretion dominated $D/Z\sim f^{\rm dep}$ has an $f^{\rm dep}$ dependence that can in principle be constrained with observations.

\subsubsection{Destruction due to Thermal Sputtering: Local Supernova Rate or Gas Temperature}

\begin{figure}
    \centering
    \includegraphics[width=\linewidth]{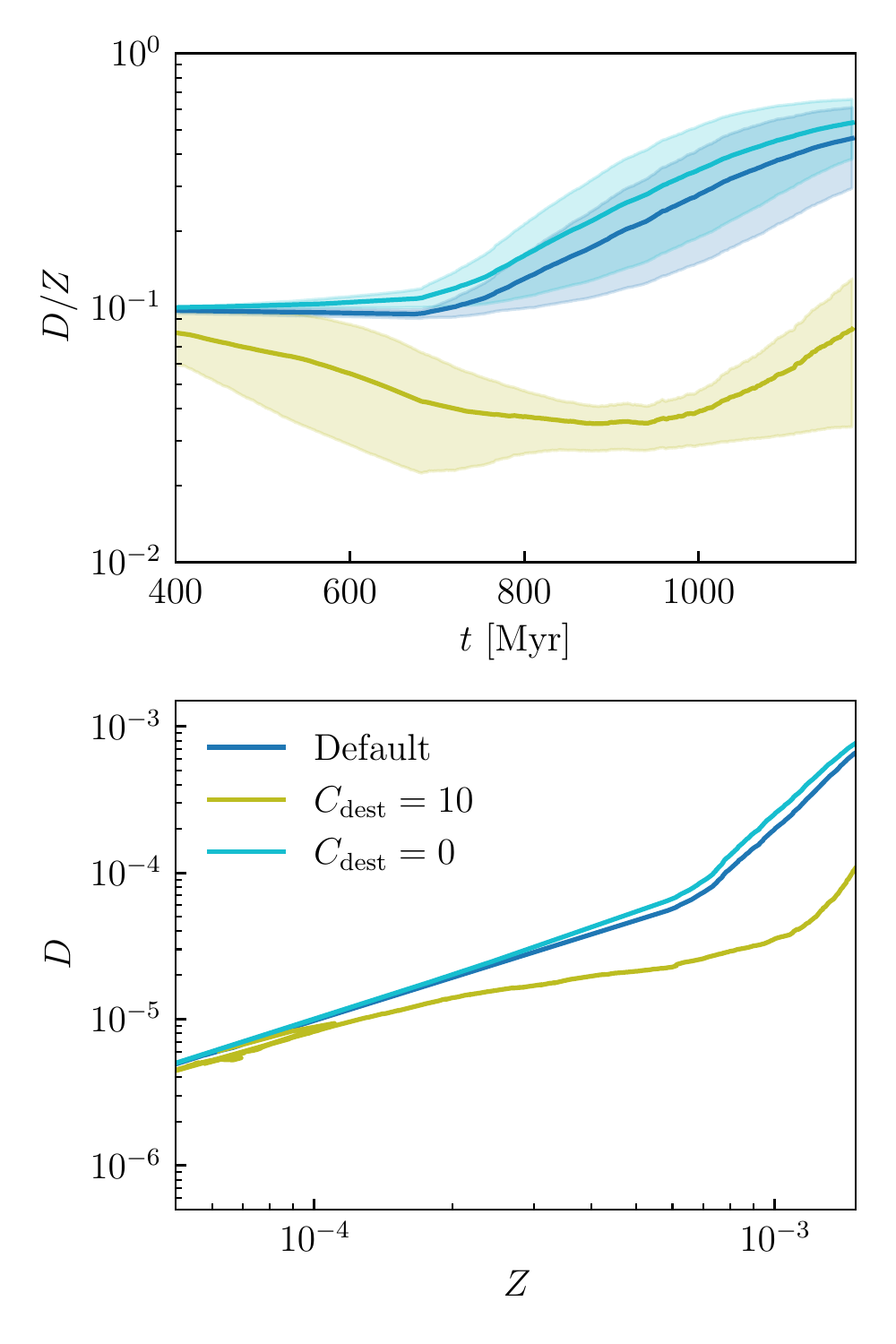}
    \caption{Effect of varying supernova remnant destruction efficiency $C_{\rm dest}$.}
    \label{fig:DtoZ_DvsZ_c_dest_SNR}
\end{figure}

\begin{figure}
    \centering
    \includegraphics[width=\linewidth]{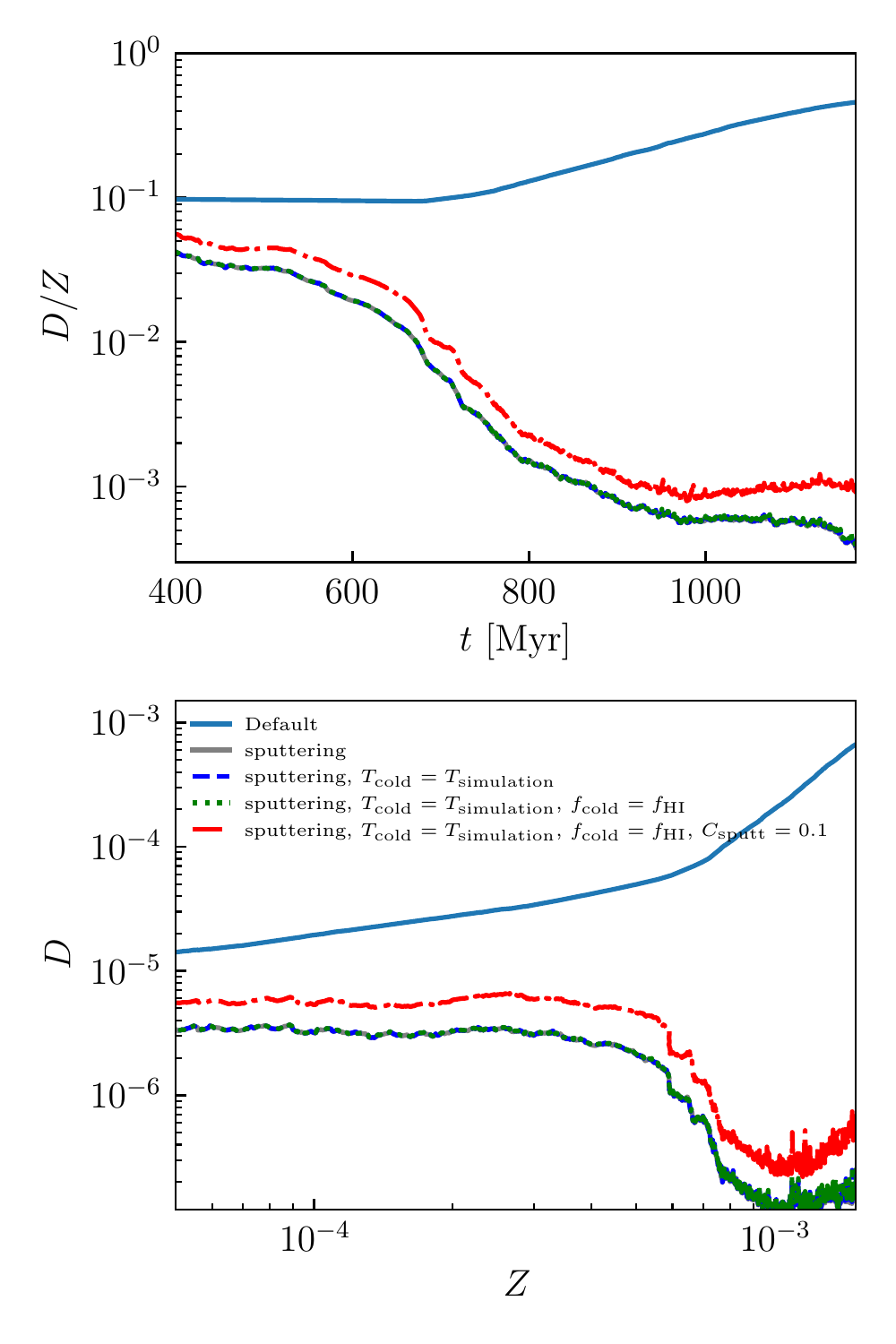}
    \caption{Effect of tying dust destruction rates to local ISM gas temperature. The grey line adopts sputtering destruction rates $\dot{D}_{\rm dest} = \dot{D}_{\rm sputt}$. This line is invisible because the dashed blue and dotted green lines -- both attempts to increase the growth rate of the dust due to ISM accretion $\dot{D}_{\rm accr}$ -- are visually indistinguishable. Thus no reasonable modification to the $\dot{D}_{\rm accr}$ overcomes the dominance of the thermal sputtering rate. This is true even if the sputtering is additionally decreased by a factor of 10, as shown with the red dot-dashed line.}
    \label{fig:DtoZ_DvsZ_sputt}
\end{figure}

Figures \ref{fig:DtoZ_DvsZ_c_dest_SNR} and \ref{fig:DtoZ_DvsZ_sputt} explore physics prescriptions and parameter choices related to the destruction rate of dust in the hot phase of the ISM due to thermal sputtering. Fig.~\ref{fig:DtoZ_DvsZ_c_dest_SNR} shows the predictions of the dust model assuming $\dot{D}_{\rm dest} = \dot{D}_{\rm SNR}$, which is our default. We display predictions for extreme variations in the efficiency of supernova remnant destruction -- one where the destruction is turned off entirely ($C_{\rm dest} = 0$) and another where it is enhanced by an order-of-magnitude $C_{\rm dest} = 10$. The $C_{\rm dest} = 0$ case corroborates the conclusion of Figure~\ref{fig:DtoZ_model_summary} that, with default parameters, supernova destruction plays a subdominant role in setting the dust content of this galaxy, as the destruction rate is always much lower than either the production or accretion rates. Therefore, turning off destruction entirely increases the dust-to-metal ratio by $\sim 10\%$ at most. However, enhancement in the SNR destruction rate by a factor of 10 is able to fully counterbalance ISM accretion and suppress $D/Z$ to values less than $y_D$ for most cosmological times in our simulation. 

Fig.\ref{fig:DtoZ_DvsZ_sputt} shows the result of an alternative choice where the destruction rate is tied explicitly to the sputtering rate predicted by the local gas density and temperature $\dot{D}_{\rm dest} = \dot{D}_{\rm sputt}$. As described in Sec.~\ref{subsubsec:dest}, this would be the more physical choice if the phase structure of the ISM were well resolved in our simulations. However, Sec.~\ref{subsubsec:ISM_phase_structure} presents evidence that this is not the case. Nonetheless, we wanted to explore the extent of the difference between a local sputtering rate prescription and the default based on local supernova rate. Fig.~\ref{fig:DtoZ_DvsZ_sputt} demonstrates that the limited resolution coupled with the delayed-cooling supernova feedback prescription predicts high ISM temperatures and therefore high sputtering rates that dominate all other processes, regardless of the choices made for accretion. Even a reduction of the sputtering rate by an order-of-magnitude does not qualitatively change this conclusion.

\subsubsection{Effect of SFR Correction}

\begin{figure}
    \centering
    \includegraphics[width=\linewidth]{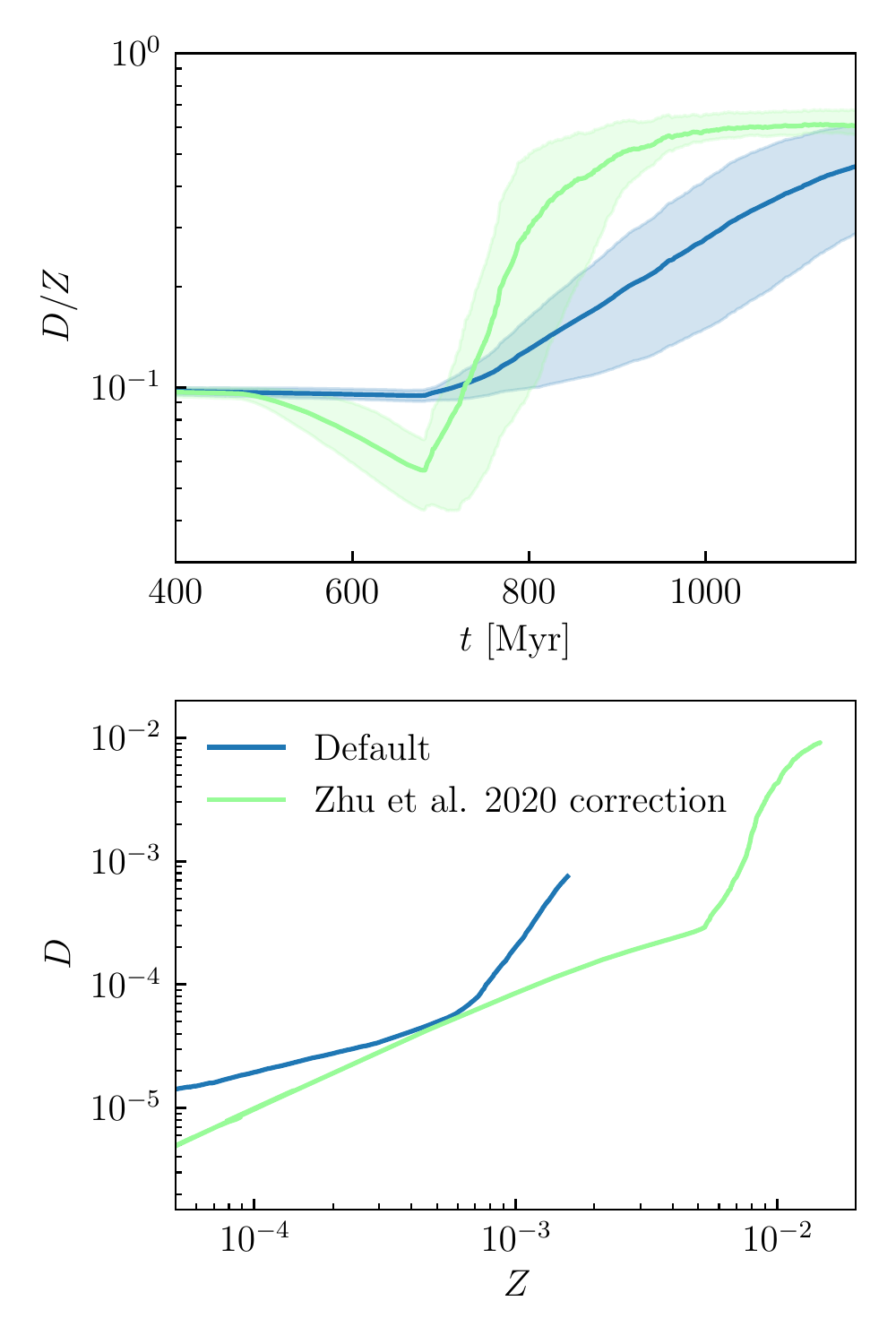}
    \caption{Effect of \citet{Zhu2020} correction for the SFR rate. $D/Z$ is initially suppressed relative to the default value because of the increased destruction rate due to supernova remnants, but much more rapidly transitions to $\sim f^{\rm dep}$ because of the enhanced ISM accretion rate due to higher metallicity.}
    \label{fig:DtoZ_DvsZ_SFR_corr}
\end{figure}

Finally we explore the effect of ``correcting'' the simulated galaxy star formation rates as would be necessary to reproduce the observed stellar-mass halo-mass relation. As described in Sec.~\ref{subsec:SFR_corr}, we multiply the star-formation-rate dependent terms -- production and supernova rates -- by a factor (Eq.~\ref{eq:SFR_corr}) that was shown in \citet{Zhu2020} to correct the CROC simulated galaxy stellar masses to approximately agree with observations. This significantly increases these rates by roughly a factor of ten by the end of the simulation. We see that initially, the enhanced star formation rate leads to an enhanced supernova destruction rate that significantly reduces the dust-to-metal ratio. However, the increased metallicity resulting from the enhanced star formation causes growth by accretion to rapidly dominate and bring $D/Z\sim f_{\rm dep}$ at earlier cosmic times. 

\subsection{Effect on Observable Quantities and Comparison to Data}

\begin{figure*}
    \centering
    \includegraphics[width=\textwidth]{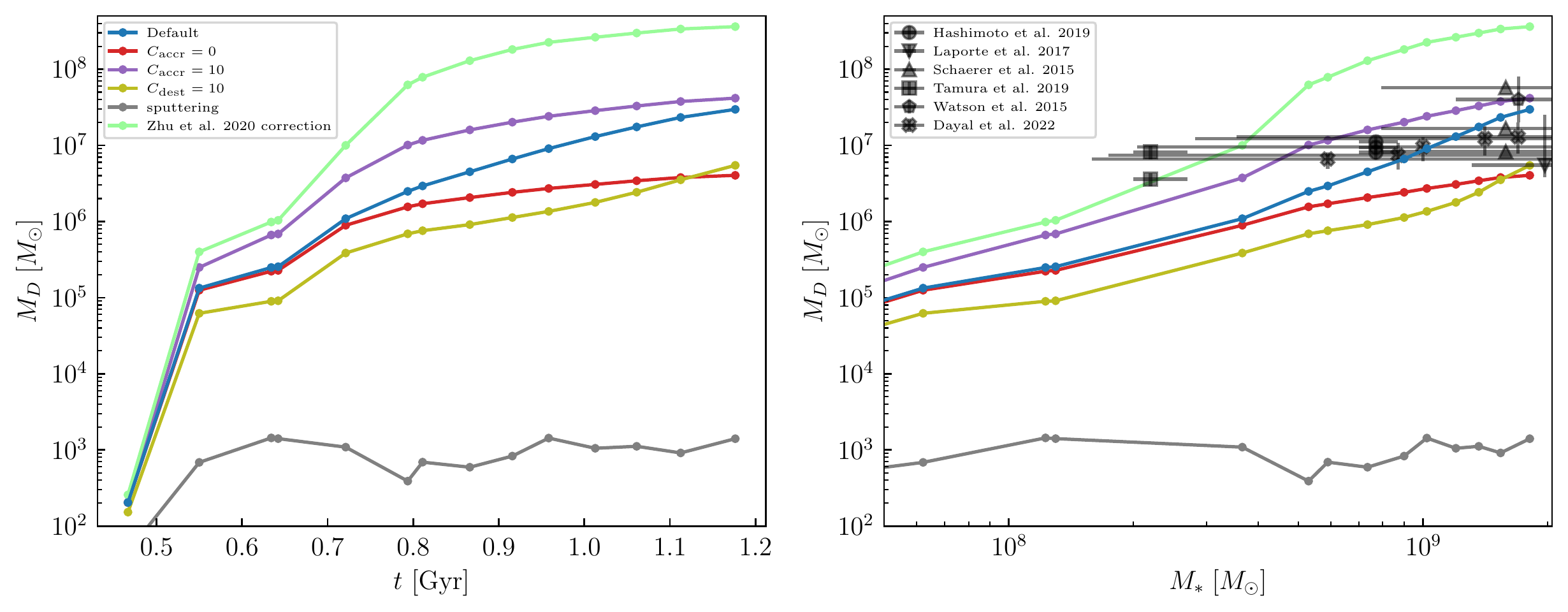}
    \caption{Dust mass for representative models as a function of cosmic time (left panel) and stellar mass (right panel). The right panel also contains observational data on the dust masses of galaxies at high redshift $z>5$ from \citet{Hashimoto2019,  Laporte2017, Schaerer2015, Tamura2019, Watson2015} and \citet{Dayal2022}.}
    \label{fig:dust_mass}
\end{figure*}

\begin{figure}
    \centering
    \includegraphics[width=\linewidth]{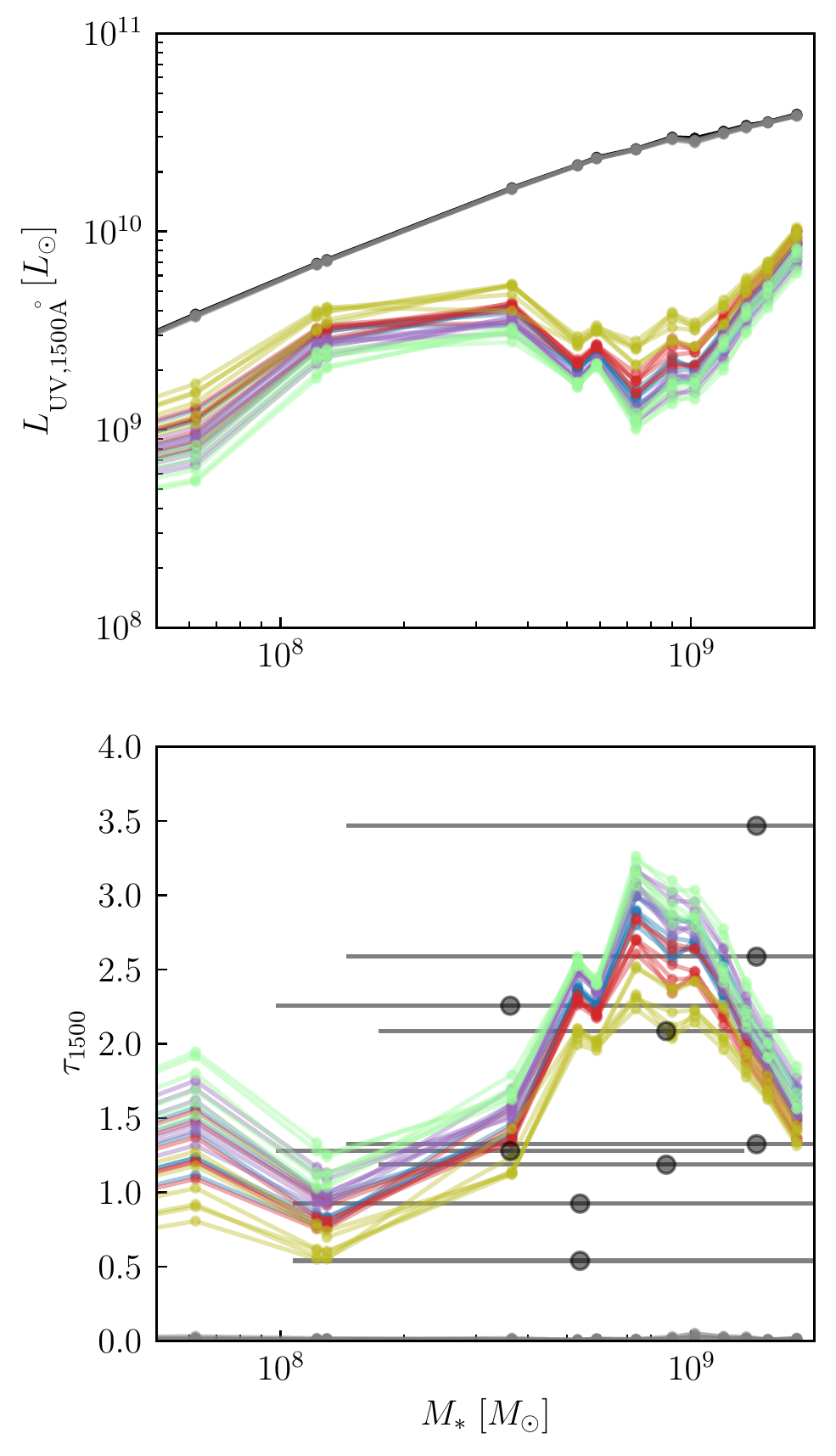}
    \caption{Ultraviolet $1500\AA$ luminosities (top panel) and optical depths (bottom panel). Data in the bottom panel are from \citet{Ferarra2022}. Colors are the same as in Figure~\ref{fig:dust_mass}.}
    \label{fig:L_UV_tau1500_Mstar}
\end{figure}

\begin{figure*}
    \centering
    \includegraphics[width=\textwidth]{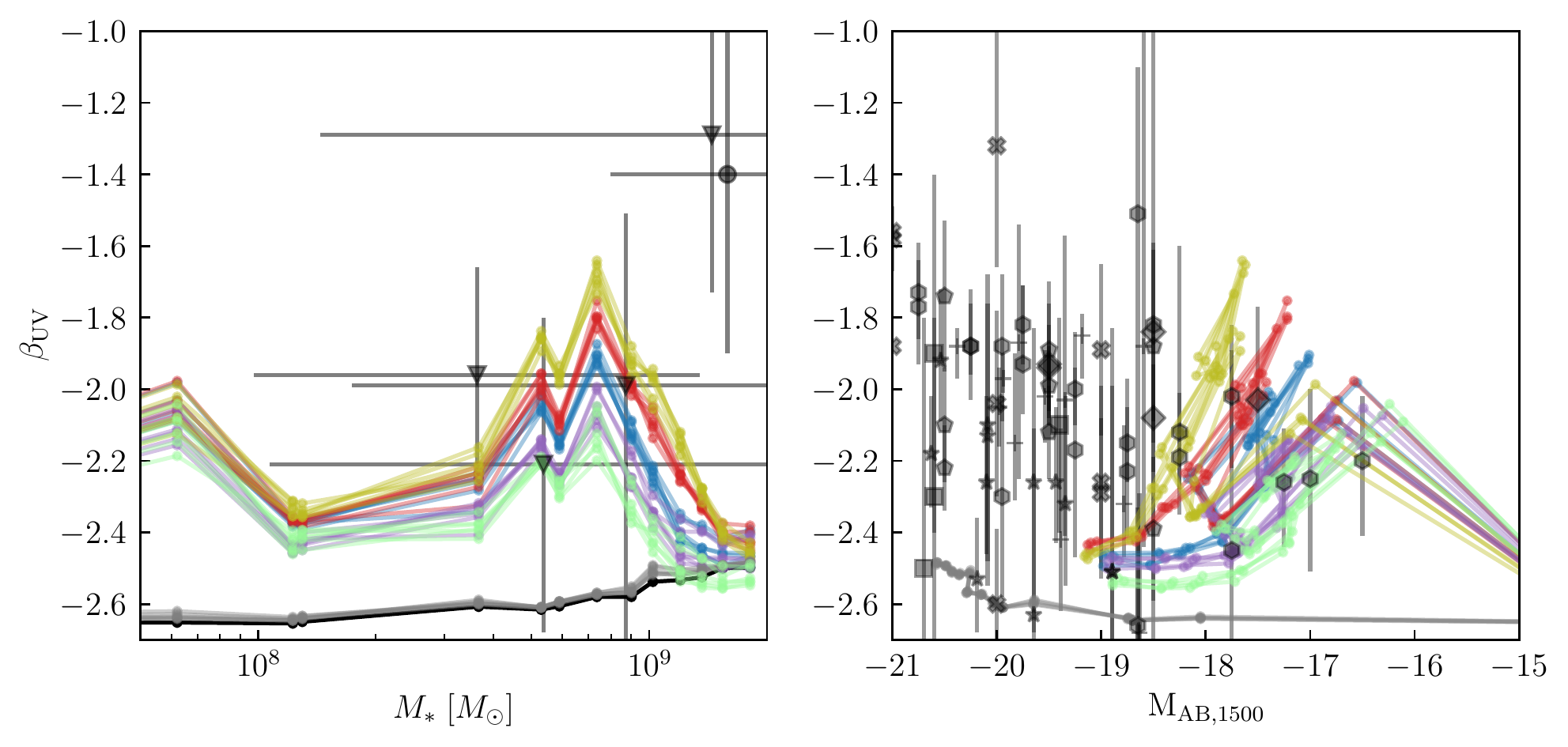}
    \caption{UV $\beta$ slope as a function of stellar mass (left panel) and UV AB absolute magnitude (right panel). Grey points are a compilation of observational data for $z>5$ galaxies from the literature: \citet[][circles]{Schaerer2015}, \citet[][downward triangles]{Ferarra2022}, \citet[][plus-signs]{Finkelstein2012}, \citet[][hexagons]{Bouwens2014}, \citet[][diamonds]{Dunlop2013}, \citet[][stars]{Bhatawdekar2021ApJ}, \citet[][filled x]{Wilkins2011}, \citet[][pentagons]{Dunlop2012}, and \citet[][squares]{Wilkins2016}.}
    \label{fig:beta_Mstar_M_UV}
\end{figure*}

Figure~\ref{fig:dust_mass} shows the total dust mass for several representative parameter choices as a function of cosmic time and stellar mass. The stellar mass panel includes observational constraints from galaxies $z\gtrsim 5$. The uncertainties in the data and a lack of a statistically interesting sample of simulated galaxies preclude a quantitative comparison, but it is encouraging that the model predicts qualitatively similar values for the dust mass of this galaxy as is suggested in observations. 

Figure ~\ref{fig:L_UV_tau1500_Mstar} shows the attenuated ultraviolet $\lambda=1500\AA$ luminosity of the simulated galaxy, and the corresponding optical depths. Colors correspond to the same dust model parameter choices as in Figure~\ref{fig:dust_mass}. Different lines of the same color indicate the same dust model observed from a different viewing angle -- six viewing angles along the positive and negative coordinate axes are calculated. Data from \cite{Ferarra2022} are shown for comparison. Figure ~\ref{fig:beta_Mstar_M_UV} shows the ultraviolet spectrum $\beta$ slope as a function of stellar mass and attenuated AB absolute magnitude at $1500\AA$, along with a compilation of $z>5$ galaxy measurements from the literature. All dust models except explicit thermal sputtering predict observations qualitatively consistent with the data. 

\begin{figure*}
    \centering
    \includegraphics[width=\linewidth]{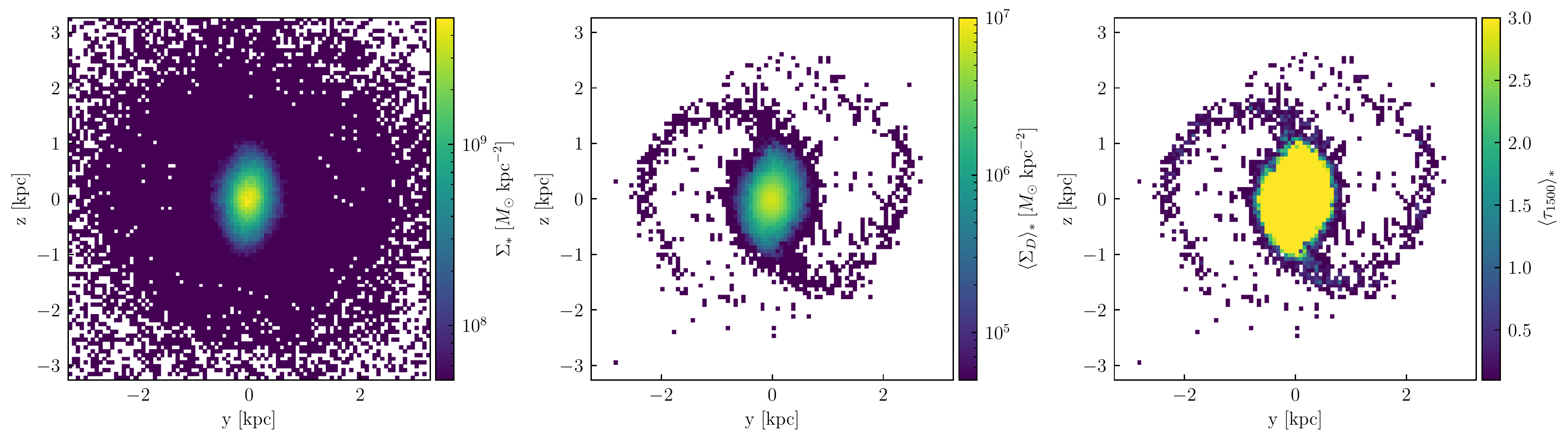}
    \caption{Maps of stellar surface density (left panel), average dust surface density (middle panel), and average stellar optical depth to dust (right panel) at $z = 5.2$. Quantities are projected along a random axis with respect to the galaxy (the coordinate z axis).}
    \label{fig:sigma_maps}
\end{figure*}

Counter-intuitively, for the non-sputtering models, the ultraviolet optical depth does not increase monotonically with increasing dust mass. This is due to the geometry of the dust distribution and consequent attenuation. Figure~\ref{fig:sigma_maps} shows the stellar mass and dust mass distributions, as well as average optical depth along the projection axis. While most of the stars in the galaxy are enshrouded in dust columns that are completely opaque, there exists an extended distribution of stars where there is effectively no dust. At late times, this component accounts for $\sim 10-30\%$ of the stellar mass, explaining the relatively low effective optical depths despite high column densities. This also explains the trend in Figure~\ref{fig:beta_Mstar_M_UV}: the extincted $\beta_{\rm UV}$ tends to the unextincted value at late times because star particles are either completely extincted due to opaque dust or they are unaffected by low dust columns.

\begin{figure}
    \centering
    \includegraphics[width=\linewidth]{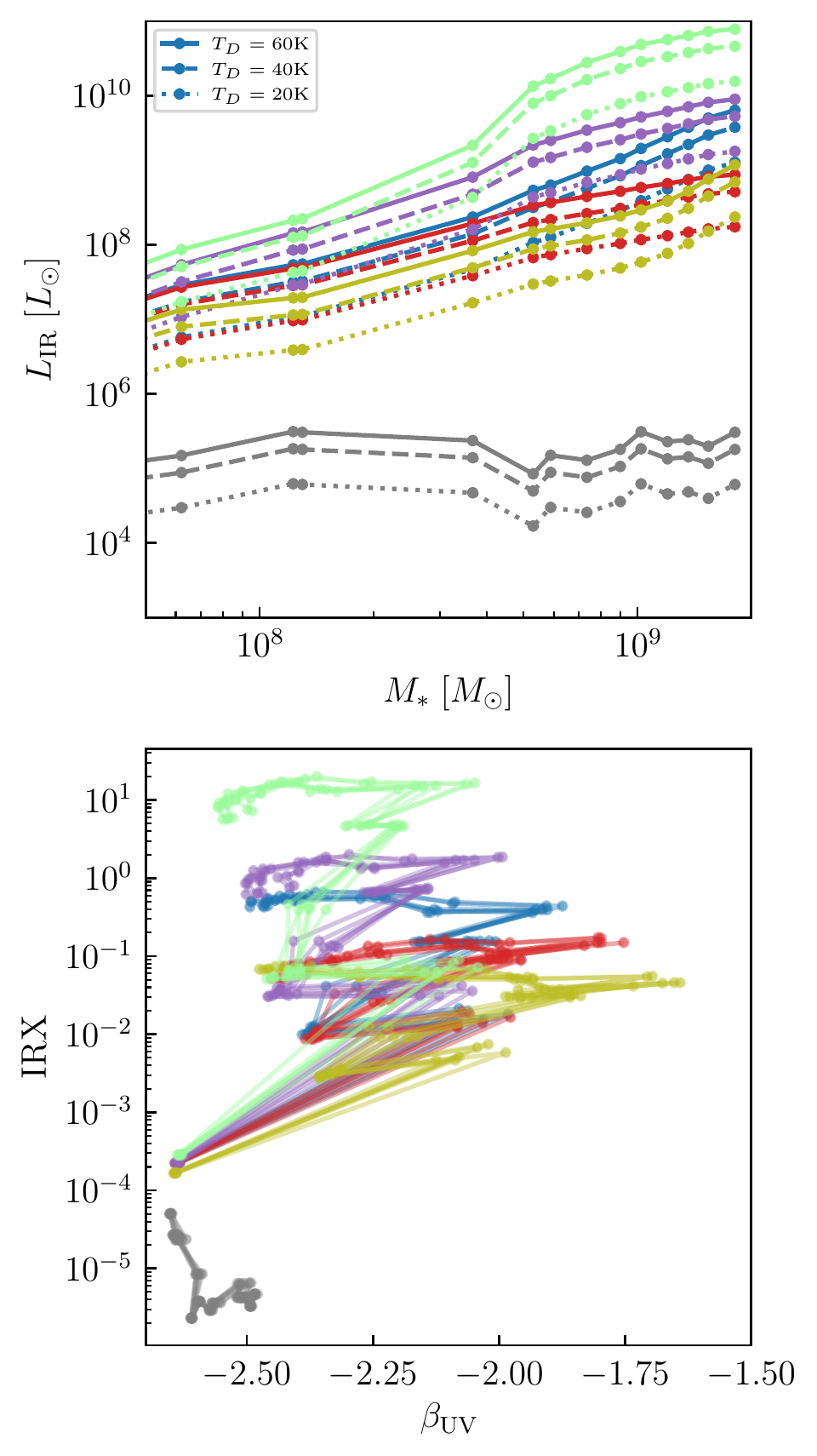}
    \caption{Infrared luminosity as a function of stellar mass (top panel) and IRX-$\beta_{\rm UV}$ relationship (bottom panel). Since we do not self-consistently calculate dust temperatures from the simulation, we show predictions for a reasonable range in the top panel. The bottom panel assumes $T_D = 40$K.}
    \label{fig:L_IR_IRX_beta}
\end{figure}

Figure ~\ref{fig:L_IR_IRX_beta} shows the infrared luminosity as a function of stellar mass for the simulated galaxy and the IRX-beta relationship. We do not attempt to self-consistently calculate the dust temperature so we present predictions for a range of reasonable values. The direct $\propto M_D$ dependence on dust mass suggests that infrared observables might be better able to distinguish between different dust models, but also highlight the necessity of accurate dust temperature measurements in simulations and observations.

\section{Discussion}

\subsection{Numerical Methods: Simulation Self-Consistency, Time-Stepping}

A major caveat to our method for predicting the dust content of simulated galaxies is the post-processing nature of our particle tracer method. In principle, metals in the dust will impact the dynamics of the ISM very differently from those in the gas-phase: dust removes metals from the gas phase that contribute to high-temperature metal-line cooling, while dust also contributes different cooling mechanisms in the form of photoelectric heating and cooling due to the emission of thermal infrared radiation. Since we post-process the tracer particles after the simulation has been run, our simulations do not take these effects into account and are therefore not entirely self-consistent. 

However, we do not think this will have a major effect for several reasons. The first is the fact that, because the phase structure of the ISM is not well resolved in these simulations and the star formation and supernova feedback prescriptions are therefore tuned \citep[see][]{Gnedin2014}, it is likely that any change in the ISM dynamics by modified cooling and heating functions could be compensated for with changes to these tuned prescriptions. Moreover, the cooling and heating functions in ISM conditions are themselves uncertain by a factor 2-3 due to the uncertainties in the rates of various cooling and heating processes \citep{Wiersma2009} and therefore excessively careful treatment of cooling and heating in simulations of this resolution is unjustified. Nonetheless, in future work we plan to implement a dust model in the simulation code that will enable self-consistency at simulation run-time, which will be crucial for higher-resolution simulations with resolved ISM dynamics.  

Time cadence is another numerical issue. The choice of $\sim 0.2$Myr interval between saved snapshots is somewhat arbitrary and is dictated by the balance between the available computational resources and the desire to fully resolve the time evolution of SN explosions from a single stellar population. We test the sensitivity of our results to this choice by down-sampling our simulation outputs by a factor of 2, i.e. removing every other output and using the remaining to calculate the dust content as described in the methods. This increases the cadence of simulation outputs to $\sim 0.4$Myr. We find that the predictions of the model under the default parameter assumptions are effectively unchanged. This is unsurprising given that average tracer quantities follow fairly smooth curves as a function of time (Fig.~\ref{fig:tracer_average_histories}). Note that this does not necessarily mean fewer timesteps were taken by the ODE solver that integrates the dust equation -- as described in the methods, this solver is allowed to take as many time-steps as necessary between simulation outputs (between which tracer quantities are linearly interpolated) based on the specified accuracy of the adaptive integration scheme. 

\subsection{Comparison to Previous Efforts and Caveats}

Several groups have recently implemented dust models similar to ours in galaxy formation simulations, either directly incorporated into the fluid-dynamical solver or in post-processing (as we do). Since this is primarily a methods paper, we will not present a comprehensive survey of these results and their comparison to ours, but instead highlight the main methodological differences between our model and others, and speculate on their effects. 

An advantage of this study compared to previous efforts is the relatively large model parameter space we have explored while simultaneously spatially resolving an individual galaxy. This is enabled by our particle tracer post-processing method, which allows us to run the full gas-dynamical cosmological galaxy formation simulation only once and subsequently post-process with a range of dust model parameters. While several studies have similarly post-processed simulations run with Lagrangian fluid dynamics solvers \citep{Mancini2015, Mancini2016, HirashitaAoyama2019, Huang2021}, only \citet{Huang2021} perform a parameter variation to explore the predictions of their model. Given the different aims of that analysis (the investigation of the full grain-size distribution in Milky Way-mass galaxies), and their use of a very different galaxy formation simulation model \citep{Weinberger2017, Pillepich2018}, our findings are complementary. This post-processing technique naturally lends itself to the inclusion of increasingly complex physics, such as an accounting for the full grain-size distribution and multiple grain species, which we plan to investigate in future work.

Figure \ref{fig:dust_mass} emphasizes how reasonable variations in uncertain parameters related to dust accretion in the ISM and destruction can lead to at least an order-of-magnitude difference in the predicted dust mass for the same galaxy. Consequently, absent a first-principles calculation or independent measurement of these parameters, the full range must be explored to honestly assess predictions in comparison to data. Of course, this also requires a statistically meaningful sample of simulated galaxies with a range of masses and formation histories to compare to the data, which we will present in forthcoming work. 

A main conclusion of our model development analysis is the importance of a carefully chosen dust destruction model. As emphasized in Figures \ref{fig:DtoZ_DvsZ_sputt} and \ref{fig:dust_mass}, the destruction rate predicted due to supernova remnants (eq.~\ref{eq:dotD_SNR}) can be vastly different from that calculated from the thermal sputtering rate expected form the erosion of grains by collisions with high-temperature gas particles (eq.~\ref{eq:sputtering}). This is likely due to the unresolevd nature of the ISM in our simulations and the choice of feedback: delayed cooing feedback appears to produce an overly-hot ISM (see Figures~\ref{fig:f_phase} and \ref{fig:n_T_dist}) but fails to drive that hot ISM out in galactic winds. The strong temperature dependence expected of the sputtering rate therefore predicts unphysically high destruction rates that prevent the accumulation of dust in the galaxy ISM to anything close to observational constraints. Previous efforts do not appear to have seen this issue because of their fundamentally different feedback model that employ a pressurized equation of state ISM and wind particles \citep{Springel2003}. In this model, the ISM gas remains at sufficiently low temperatures that thermal sputtering is never dominant and can be included along with SNR destruction without effective double-counting. This comparison demonstrates that adequate modeling of thermal sputtering is essential to correctly predicting the dust content of the hot circumgalactic and intercluter media. One hopes that a more realistically thermodynamically structured ISM would result in eq.~\ref{eq:dotD_SNR} and eq.~\ref{eq:sputtering} predicting more similar destruction rates -- as most of the hot gas that results in sputtering is expected to come from SNRs. 

\subsection{Prospects for high-$z$ Constraints}

While figures \ref{fig:DtoZ_DvsZ_yD} through \ref{fig:dust_mass} indicate the sensitivity of dust-to-metal ratio (and therefore total dust mass) on the assumed parameters of the dust model, Figures \ref{fig:L_UV_tau1500_Mstar} and \ref{fig:beta_Mstar_M_UV} indicate only modest effects on observable properties related to the produced UV radiation. This is because even the least-dust-rich models (save sputtering, which for reasons discussed above we consider unphysical), predict column densities that leave most of the central galaxy opaque. While this prediction could perhaps be sensitive to the spatial resolution of the ISM -- a more multi-phase and inhomogeneous ISM might predict a wider distribution of column densities that could have low-value tail more sensitive to the dust content -- at face-value this suggests that learning dust physics from observations that are sensitive to the rest-frame UV SED of high redshift galaxies will be inherently difficult. 

Figure ~\ref{fig:L_IR_IRX_beta} gives some hope in that the infrared luminosity is monotonically related to the total dust mass and therefore in principle more sensitive to differences between models, this observable depends even more sensitively ($\sim T^4$) on the dust temperature. On top of both of these issues is the assumed dust opacity model \citep{WeingartnerDraine2001}, which is derived from very local galaxies and therefore might be inappropriate for the high-redshift systems we investigate here. All of this suggests that future theoretical efforts will have to contend with the details of simulated ISM physics, the dust temperature, and the dust size distribution to reliably translate observational data into dust physics constraints. However, all of these statements await more conclusive judgement with a larger sample of simulated galaxies. 

\section{Conclusions}

\begin{itemize}
    \item We present a method for the prediction of the dust content of high-redshift galaxies with the Cosmic Reionization On Computers (CROC) cosmological, fluid-dynamical simulations of galaxy formation.
    \item We use a particle tracer technique to integrate an ODE for the evolution of the dust-to-gas ratio along pathlines that sample the simulated ISM in post-processing.
    \item This ODE captures a physical model for the production, growth, and destruction of dust. Dust is assumed to be produced by supernovae and AGB stars, is allowed to grow due to the accretion of metals from the gas phase of the ISM, and is assumed to be destroyed by supernovae remnant shocks. 
    \item Reproducing earlier work \citep{Genel2013}, we find that the numerical method for calculating fluid flow pathlines with particle tracers in the simulation is important forfully sampling the gas distribution in the ISM at all times in the simulation. 
    \item For simulations that do not resolve the phase structure of the ISM, different reasonable choices for grain destruction rates can predict very different predictions, especially when delayed cooling feedback is used. Delayed cooling feedback appears to produce an unrealistically warm/hot ISM, so explicitly calculating the destruction rate predicted from thermal sputtering is an over-prediction and effectively destroys all the dust. Instead, calculating the destruction rate based on the local supernova rate and assuming each remnant sweeps up a density-dependent mass of ISM, in which some fraction of dust is destroyed, produces more reasonable destruction rates. We adopt this for our default model. 
    \item Our numerical implementation reproduces the results of \citet{Feldmann2015} -- the dust-to-gas ratio passes through two regimes determined by metallicity (for a given set of model parameters): the low metallicity $D/Z$ is set by the production yields $y_D$, the high metallicity is set by the depletable fraction of metals in the ISM $f^{\rm dep}$, and the transition regime is set primarily by the timescale for grain growth in the ISM $\tau_{\rm accr}$. With default parameters, supernova are the dominant source of dust production in these early cosmic epochs, but even absent supernova dust production, AGB stars alone provide enough dust that ISM accretion can bring the dust-to-metal ratio to large values ($\sim$ tens of percent) by $z=5$.
    \item We run this dust model on the single most massive galaxy in a 10 cMpc$h^{-1}$ box ($\sim 2\times10^9 M_{\odot}$ by $z=5$). We find that the model is capable of reproducing dust masses and dust-sensitive observable quantities broadly consistent with existing data from high-redshift galaxies.
    \item The total dust mass in the simulated galaxy is somewhat sensitive to parameter choices for the dust model, since the effective growth timescale is comparable to the age of the universe at these redshifts. Consequently, observable quantities that can constrain galaxy dust mass at these epochs are potentially useful for placing constraints on dust physics in the ISM. 
    \item However, due to the geometry of the dust distribution -- which is so centrally concentrated that all viable models predict similar opacity distributions -- dust-sensitive observables due to extinction in the UV vary little with changes in dust model parameters. Moreover, the most direct measure of total dust mass -- the infrared luminosity $L_{\rm IR}$ -- is strongly sensitive to the dust temperature, which will require more careful calculations to predict self-consistently. 
\end{itemize}

\acknowledgments
This manuscript has been co-authored by Fermi Research Alliance, LLC under Contract No. DE-AC02-07CH11359 with the U.S. Department of Energy, Office of Science, Office of High Energy Physics. This work used resources of the Argonne Leadership Computing Facility, which is a DOE Office of Science User Facility supported under Contract DE-AC02-06CH11357. An award of computer time was provided by the Innovative and Novel Computational Impact on Theory and Experiment (INCITE) program. This research is also part of the Blue Waters sustained-petascale computing project, which is supported by the National Science Foundation (awards OCI-0725070 and ACI-1238993) and the state of Illinois. Blue Waters is a joint effort of the University of Illinois at Urbana-Champaign and its National Center for Supercomputing Applications. 

\bibliography{bibliography}
\end{document}